\documentclass[prb,superscriptaddress,floatfix,longbibliography,nofootinbib,letterpaper,twocolumn,aps]{revtex4}
\usepackage{natbib}
\usepackage{amsbsy}
\usepackage{scrextend}
\usepackage{graphicx}
\usepackage{color}
\usepackage[normalem]{ulem}

\usepackage[symbol]{footmisc}

\usepackage[usenames,dvipsnames]{xcolor}
\usepackage{hyperref} 
\hypersetup{%
	linktoc=page,%
	colorlinks=true,%
	linkcolor=NavyBlue,%
	citecolor=NavyBlue,%
	urlcolor=NavyBlue,%
	linkbordercolor=red,%
	pdfborderstyle={/S/U/W 1},%
}
\usepackage{amssymb,amsmath,bm,soul}
\setul{1.4pt}{.4pt}

\renewcommand{\vec}{\boldsymbol}

\newcommand{\eq}{Eq.\ }
\newcommand{\eqs}{Eqs.\ }

\newcommand{\sbonlinecite}[1]{[\onlinecite{#1}]}

\newcommand{\Tsh}{T_{\mathrm{sh}}}
\newcommand{\NS}{$\mathrm{Nb}\mathrm{Se}_{2}$}
\newcommand{\lC}{\lambda_{\mathrm{\scriptscriptstyle C}}}
\newcommand{\LC}{\Lambda_{\mathrm{\scriptscriptstyle C}}}

\begin{document}
\title{{
Thermal hysteresis of the Campbell response as a probe for bulk pinscape spectroscopy
}}
\author{Roland Willa}
\affiliation{
        Materials Science Division,
	    Argonne National Laboratory,
	    Lemont IL, United States of America}
\author{Mariano Marziali Berm\'udez}
\affiliation{Universidad de Buenos Aires, Facultad de Ciencias Exactas y Naturales, Departamento de F\'isica, Buenos Aires, Argentina}

\affiliation{CONICET-Universidad de Buenos Aires, Instituto de F\'isica de Buenos Aires, Buenos Aires, Argentina}

\affiliation{Instituto de F\'isica de L\'iquidos y Sistemas Biol\'ogicos (IFLYSIB), UNLP-CONICET, La Plata, Argentina}
\author{Gabriela Pasquini}
\affiliation{Universidad de Buenos Aires, Facultad de Ciencias Exactas y Naturales, Departamento de F\'isica, Buenos Aires, Argentina}

\affiliation{CONICET-Universidad de Buenos Aires, Instituto de F\'isica de Buenos Aires, Buenos Aires, Argentina}
\date{\today}

\begin{abstract}
In type-II superconductors, the macroscopic response of vortex matter to an external perturbation depends on the local interaction of flux lines with the pinning landscape (pinscape). The (Campbell) penetration depth $\lC$ of an $ac$ field perturbation is often associated with a phenomenological pinning curvature. However, this basic approach is unable to capture thermal hysteresis effects observed in a variety of superconductors. The recently developed framework of strong-pinning theory has established a quantitative relationship between the microscopic pinscape and macroscopic observables. Specifically, it identifies history-dependent vortex arrangements as the primary source for thermal hysteresis in the Campbell response. In this work, we show that this interpretation is well suited to capture the experimental results of the clean superconductor {\NS}; as observed through Campbell response (linear $ac$ susceptibility) and small-angle neutron scattering measurements. Furthermore, we exploit the hysteretic Campbell response upon thermal cycling to extract the temperature dependence of microscopic pinning parameters from bulk measurements; specifically the pinning force and pinning length. This spectroscopic tool may stimulate further pinscape characterization in other superconducting systems.
\end{abstract}

\maketitle

\section{Introduction}\label{}
In type-II superconductors, magnetic field threads the material in form of lines, so-called vortices, with quantized magnetic flux $\Phi_{0} \!=\! hc/2e$. In a (meta)stable vortex state the pinning forces exerted by the distribution of material impurities is balanced by elastic forces due to vortex-vortex interactions, which, in turn, depend on the vortex arrangement. Pinned by these material defects, vortices resist to motion upon applying a subcritical current $j \!<\! j_{c}$ and thereby allow for dissipation-free transport of electrical current. The response of a pinned vortex state therefore depends both on the material heterogeneity (pinning landscape) and on the specific flux-line distribution. Tailoring material properties for specific high-power applications therefore involves to optimize the subcritical supercurrent and to minimize effects of thermally activated motion (creep); a task that is undertaken by joining analytical, numerical, and experimental efforts\cite{Blatter1994, MacManus2004, Haugan2004, Kang2006, Silhanek2007, Gutierrez2007, Maiorov2009, Polat2011, Miura2013b, Ray2013, Kwok2016, Sadovskyy2016b, LeThien2017, Eley2018a}. A quantitative understanding of the pinning problem will provide vital input for next-generation superconducting technology. 

The description of a pinned vortex state and its macroscopic response to external perturbations based on the microscopic interactions with defects poses a complex problem. Quantitative insight can be gained when the superconducting matrix contains a dilute distribution of defects. In this limit, a strong-pinning framework, originally proposed by Labusch \cite{Labusch1969} and Larkin and Ovchinnikov \cite{Larkin1979}, has recently been developed to provide quantitative expressions for macroscopic observables, such as the critical current\cite{Blatter2004} $ j_{c}$, the excess current characteristic \cite{Thomann2012, Thomann2017, Buchacek2018a}, and the Campbell penetration depth\cite{Willa2015a,Willa2015b,Willa2016} $ \lC$ of small-field oscillations.

The direct link between the microscopic pinning configurations of a vortex state and its response to bulk measurements invites one to follow the inverse route, namely, to distill specific the pinscape characteristics from experimental data. However, the fact that a combination of different microscopic ingredients enters the final response often obstructs this path. In a recent work\cite{Willa2015a}, the hysteresis in the Campbell penetration (or $ac$ susceptibility) measurement has been identified as a potential tool to circumvent this obstruction. Specifically, it has been shown that different microscopic realizations of the pinned vortex state result in different (Campbell) responses. Tuning the former and recording the latter provides different viewpoints on the same pinscape. In the present work, we push this idea further and develop a formal set of rules characterizing the vortex state's evolution upon changing temperature or applying a large $ac$ field oscillations. Reversing the argument, we then discuss an experimental protocol of the Campbell response that allows to probe the temperature evolution of microscopic pinning parameters.

The necessary tuning knobs---temperature and $ac$ field shaking---to exploit this spectroscopic tool have recently been used \cite{Pasquini2008, Perez2011, Marziali2015, Marziali2017} to investigate the linear $ac$ susceptibility of {\NS} single crystals; a superconductor with a scarce distribution of structural defects%
\footnote{%
The in-plane distance between atomic defects is much larger than the superconducting coherence length and, for typical magnetic fields ($\sim $ kOe), it is even larger than the inter-vortex distance. The surface-density of atomic defects on a growth surface is generally below $10^{-4} \mathrm{nm}^{-2}$. The density of topological defects observed by STM are even lower, see Ref.\ \sbonlinecite{Guillamon2009-thesis}.}
and where dynamic and thermal effects in vortex matter have been studied extensively\cite{Henderson1998,Xiao2004,Li2006,Yaron1995,Pasquini2008,Perez2011, Marziali2015, Marziali2017}.
In Refs.\ \sbonlinecite{Marziali2015, Marziali2017}, joint small-angle neutron scattering (SANS) and in-situ $ac$ susceptibility measurements have been used to investigate the vortex lattice structure and its response to $ac$ magnetic perturbations. Various cooling and warming protocols have revealed a strong hysteresis in the $ac$ response of this material. In addition to measurements in the linear response regime, the setup has been operated at large $ac$ field amplitudes to shake the vortex state, leading to a dynamic reorganization. At low temperatures, after shaking, one observes a significant change in the linear $ac$ response and a increased flux-line lattice order. Both effects have commonly been attributed to the reduction of dislocations in the vortex lattice. Applying cooling and warming protocols without dynamic assistance reveal a strong hysteresis in the linear response of this material, which, however, does not correlated with a measurable change in the mean dislocation density. The dislocation-based picture is therefore insufficient to explain these observations.

Whereas a qualitative interpretation of the experimental findings within a dislocation-based picture has been proposed \cite{Marziali2017}, it does not fully account for all experimental observations.
In the following, we shall provide a new interpretation of these experiments, based on the strong-pinning formalism, where the hysteresis in the $ac$ linear response originates from different microscopic realizations of the pinned vortex state. Moreover, we use the recorded data to test the proposed spectroscopic tool for investigating microscopic pinning parameters. Here, the term pinscape spectroscopy refers to the task of deconvoluting the thermal evolution of bulk properties to access microscopic quantities; in this particular case, the pinning force and length in {\NS}. This concept substantially differs from spectroscopic methods associated with local probe techniques, which allow to study surface properties of single crystals \cite{Hess1989, Raes2014, Timmermans2014, Guillamon2008b, Galvis2017} or films \cite{Embon2015}.

This paper is organized as follows: 
We revisit in section \ref{sec:theory} the key ingredients to describe the Campbell response in the framework of strong vortex pinning and derive a set of rules governing its evolution upon changing temperature or applying a shaking field.
In section \ref{sec:hysteresis}, we summarize experimental studies on {\NS} \sbonlinecite{Pasquini2008, Perez2011, Marziali2015, Marziali2017} and interpret the findings within the strong-pinning picture. 
In section \ref{sec:extraction}, we show how the hysteresis in the Campbell response---as caused by thermal cycling---may be used to extract the temperature evolution of characteristic pinned parameters (force and length). 
This protocol is applied to $ac$-susceptibility data on {\NS}.

\section{Response to weak and strong ac fields within strong-pinning theory}\label{sec:theory}

\begin{figure}[t]
\centering
\includegraphics[width = .42\textwidth]{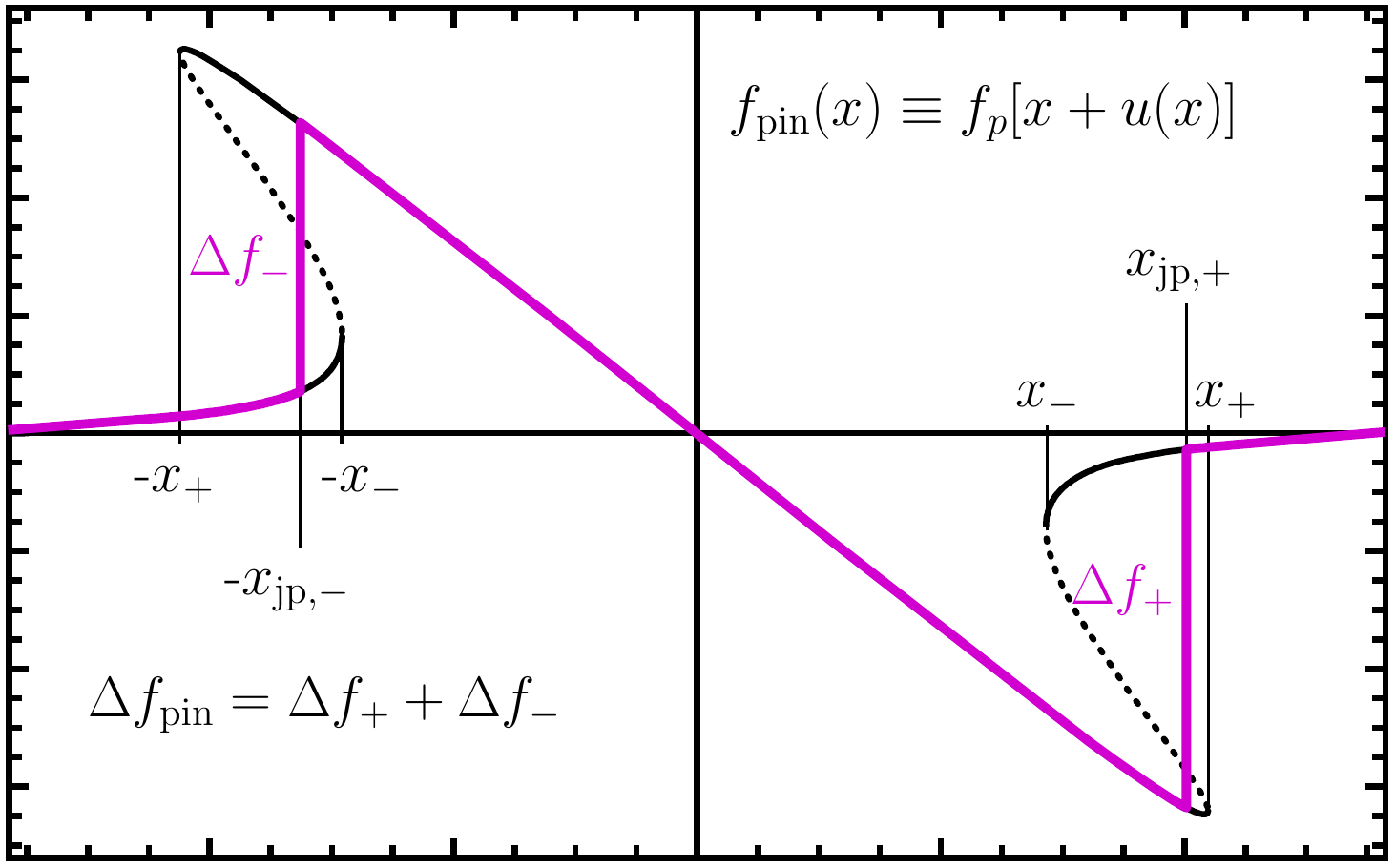}
\caption{
Multivalued pinning force $f_{\mathrm{pin}}(x)$ caused by an elastic flux-line interacting with a strong defect. 
The branch occupation (purple) depends on the state preparation and features discontinuities at $\pm x_{\mathrm{jp},\pm}$. 
These pinning length are limited by the extremal values $x_{\pm}$ marking a softening of the vortex deformation.
The sum $\Delta f_{\mathrm{pin}}$ of the associated force jumps $\Delta f_{\pm}$ determines the Campbell length $\lC$ as it enters the on average potential curvature, see \eqs \eqref{eq:Campbell-length} and \eqref{eq:alpha-sp}.}
\label{fig:fpin}
\end{figure}

A brief account on the $ac$ response of a pinned vortex state within the strong-pinning paradigm shall be given here, while more details are found in Ref.\ \sbonlinecite{Willa2015a, Willa2015b, Willa2016}. In this context, the relevant regimes are (i) the non-dissipative and low-amplitude (Campbell \cite{Campbell1969, Campbell1971}) regime and (ii) the non-linear large-amplitude (Bean \cite{Bean1962}) regime with dynamic reordering of the vortex state, see also Ref.~\sbonlinecite{vanderBeek1993}. Typically, the $ac$ response is formulated through the susceptibility $\chi$ or the (Campbell or Bean) penetration depth $\lambda$. Both quantities express the material's shielding capability and---if $\lambda$ is small compared to the typical sample size $R$---are approximately related by $\lambda \approx (1 + \chi) R$; a relation we will use in the following.

\subsection{Theory of strong vortex pinning}\label{sec:spt}
The theory of strong vortex pinning applies to superconductors with a low defect density $n_{p}$, i.e., where each defect acts independently and the resulting pinning force density is directly proportional to $n_{p}$. The bare pinning force $\vec{f}_{\!p}(\vec{r})$ exerted by a point-like defect on a stiff vortex at a distance $|\vec{r}|$ turns into a multivalued force $\vec{f}_{\!\mathrm{pin}}(\vec{r}) \equiv \vec{f}_{\!p}(\vec{r}+\vec{u})$ once the vortex is free to bend. The microscopic displacement $\vec{u}$ of the vortex at the height of the defect satisfies the condition 
\begin{align}\label{nonlinear}
\vec{f}_{\!p}(\vec{r}+\vec{u})=\bar{C}\vec{u},
\end{align}
i.e., a microscopic force balance equation between the pinning force $\vec{f}_{\!p}(\vec{r}+\vec{u})$ exerted on the vortex and the restoring force $\bar{C}\vec{u}$ acting on the deformed vortex. Here, the lattice elasticity $\bar{C}=[\int d^{3}k/(2\pi )^{3}G_{xx}(\vec{k})]^{-1}$ involves  \cite{Blatter2004} the elastic Green's function $G_{\alpha \beta }(\vec{k})$ of the flux-line lattice. Equation $\eqref{nonlinear}$ features multiple solutions $\vec{u}(\vec{r})$ when the Labusch parameter \cite{Labusch1969} $\kappa \equiv \max (f_{p}^{\prime })/\bar{C}$ is larger than unity. The criterion $\kappa > 1$ defines the strong-pinning regime where vortices placed at distances $|\vec{r}| \in [x_{-},x_{+}]$ from the defect can realize one of two stable solutions: a pinned $\vec{u}_{p}$ and a free one $\vec{u}_{f}$, see Fig.~\ref{fig:fpin}. 
The boundaries $|\vec{r}| = x_{\pm}$ appear as instabilities in the solution of \eq \eqref{nonlinear}, i.e.,  $du(r)/dr|_{r=x_\pm} \to \infty$.

A random distribution of sparse defects implies that each vortex-to-pin separation $\vec{r}$ is realized with equal probability. In the case of strong pinning, $\kappa \!>\! 1$, the ambiguity of populating the branches in the multivalued region is resolved by studying the preparation of the vortex state. In the following, we consider a simple geometry, where vortices---directed along $z$---are driven along a fixed direction $x$. Then, the problem can be decomposed into a longitudinal part (along $x$) and a transverse direction (along $y$).
A general pinned vortex state can be described by two longitudinal positions $-x_{\mathrm{jp},-} < 0$ and $x_{\mathrm{jp},+} > 0$---where the occupation changes from the unpinned to the pinned branch and back, see Fig.\ \ref{fig:fpin}---and one transverse trapping radius $r_{\perp}$. For simplicity, we choose $x_{\mathrm{jp},\pm}$ to lie in the (positive) interval $[x_{-}, x_{+}]$, and consider all vortices within the transverse impact distance $|y| \!<\! r_{\perp}$ equivalent to the case $y \!=\! 0$.

If the magnetic field is turned on in the low-temperature, superconducting phase, vortices penetrate the sample from the surface, establishing a critical (or Bean) profile. The field gradient of this zero-field cooled (ZFC) state breaks the symmetry between the penetration direction ($x$) and the direction ($y$) transverse to it.
Along $x$, the vortex state features a very asymmetric branch occupation with the pinned branch occupied between $-x_{-}$ and $x_{+}$ (i.e., $x_{\mathrm{jp},\pm} = x_\pm$).
In the transverse direction $y$, only vortices within the defect's trapping radius $r_{\perp} \!\approx\! x_{-}$ get pinned.

In contrast, when the magnetic field is turned on in the normal phase, and later cooled through the superconducting transition (field-cooled or FC state), the vortex density features no field gradient. The absence of net currents implies a vanishing net pinning force and a rotationally symmetric branch occupation. Within the simplifying decomposition in longitudinal and transverse components this implies $x_{\mathrm{jp},+}\!=\!x_{\mathrm{jp},-}\!=\!r_{\perp} \equiv x_{\mathrm{jp},0}$, i.e., vortices satisfying $|x|, |y| < x_{\mathrm{jp},0}$ realize the pinned solution $\vec{u}_{p}$. Vortices further away will barely be affected by the defect and realize the free solution $\vec{u}_{f}$. For a detailed discussion on the position $x_{\mathrm{jp},0}$ and its temperature dependence, the reader is referred to Ref.~[\onlinecite{Willa2016}].

It is the microscopic realization of pinning states that dictates the system's macroscopic response to an external perturbation. For the particular case of the $ac$ susceptibility, this dependence shall be discussed in the following.

\subsection{Campbell regime - Linear response}\label{sec:Campbell}
\begin{figure}[t]
\centering
\includegraphics[width = .38\textwidth]{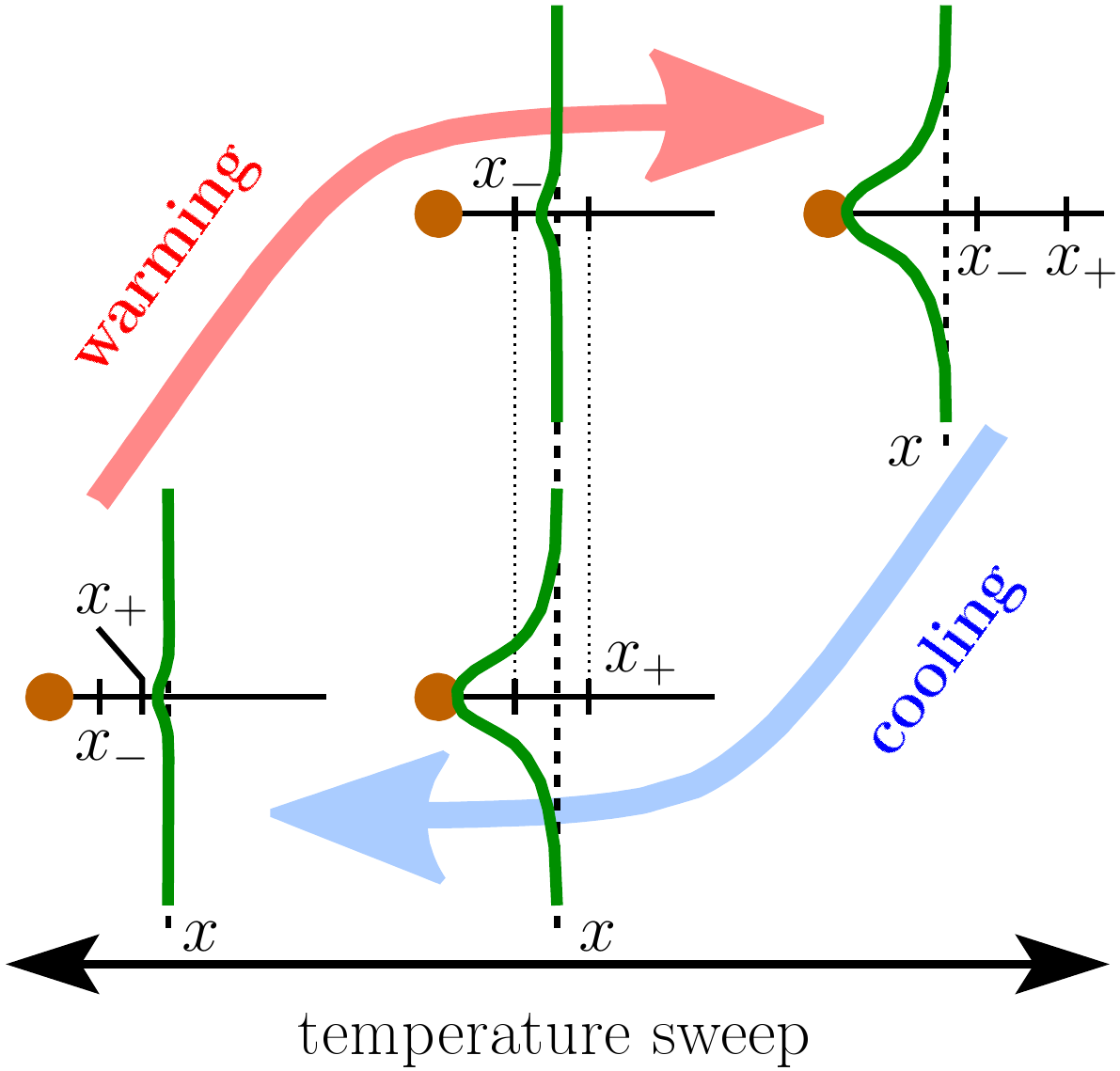}
\caption{
The origin of hysteresis in the Campbell response, illustrated by the pinning state of a single vortex at a distance $x$ from the defect: As temperature is decreased, $x_{+}(T)$ moves through $x$, and the vortex undergoes a depinning instability (lower sketches from right to left). When temperature is subsequently increased, $x_{-}(T)$ moves through the position $x$ of the unpinned vortex, which now gets trapped by the defect (upper sketches from left to right).
No instability takes place in the other two situations, i.e., when $x_{+}$ ($x_{-}$) passes through $x$ for an unpinned (pinned) vortex.
In the intermediate temperature window, the vortex remains pinned during cooling and unpinned during warming, resulting in different Campbell responses.
}
\label{fig:illustration}
\end{figure}
For many years, the response of the vortex lattice to a weak perturbation by an $ac$ magnetic field has been described by a phenomenological theory where, for small vortex displacements $U$ in the neighborhood of their equilibrium positions, the pinning potential  is modelled by an harmonic well $\alpha U^{2}$.  This results in a linear restoring force $F_{\mathrm{pin}}(U)=-\alpha U$ for which\cite{Campbell1969, Campbell1971} the $ac$ perturbation decays within the sample on the length $\lC(\omega )=[B^{2}/4\pi (\alpha -i\eta \omega)]^{1/2}$, with $B$ the field strength, $\eta $ the Bardeen-Stephen viscosity, and $\omega $ the
$ac$ drive frequency [this frequency is assumed to be large in comparison with the timescale\cite{Brandt1994a} of (thermal) creep $\tau_{c}$, i.e. $\omega \tau_{c} \gg 1$]. At low frequencies, $\eta \omega \ll \alpha $, this length reduces to the frequency-independent Campbell penetration depth
\begin{align}\label{eq:Campbell-length}
\lC = (B^{2}/4\pi \alpha)^{1/2}.
\end{align}

From a microscopic treatment of the pinning problem within the strong-pinning formalism \cite{Blatter2004}, it is known that the overall restoring force on the vortex lattice results from proper averaging of pinning forces acting on individual vortices. Conceptually the same applies to the harmonic response of pinned vortices in the Campbell regime\cite{Willa2015a, Willa2015b, Willa2016}, where a small displacement $U$ of the vortex system results in a linear restoring force $F_{\mathrm{pin}}(U)=-\alpha _{\mathrm{sp}}U$, with 
\begin{align}\label{eq:alpha-sp}
\alpha _{\mathrm{sp}}=2 n_{p} r_{\perp }\Delta f_{\mathrm{pin}}/a_{0}^{2}
\end{align}
an (averaged) strong-pinning curvature\cite{Willa2016}. This \textit{spring constant} scales linearly with the density of defects $n_{p}$, the (transverse) trapping radius $r_{\perp}$, and the jump in the (longitudinal) force profile, denoted by $\Delta f_{\mathrm{pin}}$. Here, $a_{0} = (\Phi _{0}/B)^{1/2}$ is the inter-vortex distance and $\Delta f_{\mathrm{pin}} = \Delta f_{+} + \Delta f_{-}$ measures the sum of the force discontinuities in the multivalued force at $\pm x_{\mathrm{jp},\pm }$, see Fig.\ \ref{fig:fpin}.

Upon warming or cooling the system, the extremal distances $x_{\pm}$ of the pinscape---determining the range where bistable solutions are allowed---evolve. 
As a result, a vortex initially lying within the interval $[x_{-}(T),x_{+}(T)]$ may fall outside the interval $[x_{-}(T+\delta T),x_{+}(T+\delta T)]$ and vice versa, thus switching its pinning state. 
More generally, the description of the microscopic occupation of the force branches, and subsequently of the Campbell length, require to study the evolution of the distances $x_{\mathrm{jp},\pm}$ and $r_{\perp }$ with temperature.

Thermal evolution can be expressed as the set of rules discussed below. A related discussion of the branch evolution near the Labusch point $\kappa = 1$ is provided in Ref.\ \sbonlinecite{Willa2016}.
In order to simplify the following discussion, we will use the symbol $\ell$ to refer to any of the lengths associated with the force jump positions ($x_{\mathrm{jp},+}$, $x_{\mathrm{jp},-}$, or $r_{\perp}$).

\textbf{Case \emph{a} - }
When the discontinuities in branch occupation at $\ell \in \{ x_{\mathrm{jp},+}, x_{\mathrm{jp},-}, r_{\perp}\}$  are away from the extremal points $x_{\pm}$, such as depicted in Fig.\ \ref{fig:fpin}, a small change (positive or negative) in temperature, $T \to T \!+\! \delta T$, does not force any instability. Therefore, vortices preserve their initial state, whether pinned or unpinned. In this case, we have
\begin{align}
\ell(T\!+\!\delta T) = \ell(T).
\end{align}
This implies that only the pinning force profile $f_p$ evolves with temperature, while branch occupation remains unchanged.
This case is fully reversible and no hysteresis is expected for the associated Campbell response. 
A particular realization of this case has been discussed in Ref.\ \sbonlinecite{Willa2016} when going through the Labusch (or weak-to-strong pinning transition) point.

\textbf{Case \emph{b} - }
Branch occupation changes with temperature when vortices are forced to switch from an unpinned to a pinned state or vice versa because of the disappearance of the former solution.
This situation occurs in two flavors:
On the one hand, an originally unpinned vortex at $|x| \!=\! x_{-}$ will get pinned if $x_{-}(T)$ \emph{increases} upon a temperature change and remain unpinned otherwise. The evolution of any length  $\ell(T)$ initially matching $x_{-}(T)$ is therefore dictated by the asymmetric rule
\begin{align}\label{eq:caseb}
\!\!\!\ell (T\!+\!\delta T)=\left\{
\begin{aligned}
&x_{-}(T) &&\!\text{for\ } x_{-}(T\!+\!\delta T) < x_{-}(T) \\
&x_{-}(T\!+\!\delta T) &&\!\text{for\ } x_{-}(T\!+\!\delta T) > x_{-}(T).\!
\end{aligned}\right.
\end{align}
This situation is expected when the ratio $f_{p} / \bar{C}$ between the pinning force and the effective elasticity increases.
On the other hand,
an originally pinned vortex at $|x| \!=\! x_{+}$ will depin from the defect if $x_{+}(T)$ \emph{decreases} upon a temperature change and remain pinned otherwise. Hence, for $\ell(T)$ initially matching $x_{+}(T)$ we have
\begin{align}\label{eq:casebprime}
\!\!\!\ell (T\!+\!\delta T)=\left\{
\begin{aligned} &x_{+}(T\!+\!\delta T) &&\!\text{for\ } x_{+}(T\!+\!\delta T) < x_{+}(T)\\
&x_{+}(T) &&\!\text{for\ } x_{+}(T\!+\!\delta T) > x_{+}(T).\!
\end{aligned}\right.
\end{align}
This situation is expected when the ratio $f_{p} / \bar{C}$ between the pinning force and the effective elasticity decreases.

The hysteresis in the Campbell response when applying a thermal cycling protocol is rooted in the asymmetry of this case \emph{b}. 
Indeed, when the length $\ell$ is forced by rule \emph{b} to follow the evolution of $x_{\pm }$ upon cooling, it will not do so upon warming (where, instead, the system will follow case \emph{a}) and vice versa.
For illustration purposes, Fig.\ \ref{fig:illustration} depicts the hysteresis effect for a single vortex at a distance $x$ as the critical lengths $x_{\pm}$ move past $x$ upon changing the system's temperature.

\subsection{Bean regime - Non-linear response}

In a critical state, the Lorentz force density $F_{\scriptscriptstyle L}=j_{c}B/c$ (cgs) is counterbalanced by the maximal pinning force density provided by the pinscape. When shaking the vortex system with a large $ac$ field at low frequencies, displacements much larger than the typical pinning length $x_{+}$ are imposed on the vortices and the local currents reach the critical value $\pm j_{c}$. 
As a consequence, a critical profile is established and the vortex density gradient is fixed to $\partial_{x} B_{z} = \mp (4\pi /c)j_{c}$. 
This critical state penetrates to a (Bean) depth 
\begin{align}
\lambda_{\mathrm{\scriptscriptstyle B}} = ch_\mathrm{ac} / 4\pi j_{c},
\end{align}%
which depends linearly on the amplitude of the $ac$ field and hence the $ac$ magnetic response is non-linear. As a function of the amplitude $h_\mathrm{ac}$ of the $ac$ field, the response crosses over from the linear Campbell regime ($h_\mathrm{ac}\ll j_{c}\lambda _{\mathrm{\scriptscriptstyle C}}/c$) to the non-linear Bean regime \cite{Willa2015b} when $h_\mathrm{ac} \sim j_{c}\lC/c$.
Upon turning the shaking field off, the system gets locked in a final state which depends on the damping time scale of the shaking field.

\textbf{Case \emph{c} -}
Whereas different ramping velocities are discussed in Appendix \ref{app:ramping}, a rapid turn-off installs a Bean critical state (similar to the ZFC case discussed above) where the branch occupation is maximally asymmetric%
\footnote{
If the shaking field is turned off adiabatically slowly, rule \emph{c} is replaced by $x_{\mathrm{jp},\pm }\rightarrow [ x_{+}(\Tsh)+x_{-}(\Tsh)]/2$, and $ r_{\perp }\rightarrow x_{-}(\Tsh)$.
}%
:
\begin{align}
\!x_{\mathrm{jp},\pm}\!\rightarrow x_{\pm}(\Tsh) \quad \text{and} \quad r_{\perp }\rightarrow x_{-}(\Tsh).
\end{align}%
Note that this rule is independent of the vortex state prior to shaking:\ the large-amplitude field oscillations determine the final state, erasing the system's memory.

\section{History effects in $\mathbf{NbSe}_{2}$}\label{sec:hysteresis}
Equipped with a formalism to describe the linear Campbell and non-linear Bean dynamics within the strong-pinning paradigm, we proceed with a comparison of the experiments on {\NS} and provide support for an interpretation within the microscopic picture of strong vortex pinning.

\subsection{Experimental evidence}\label{sec:experiments}
\begin{figure}[t]
\centering
\includegraphics[width = .44\textwidth]{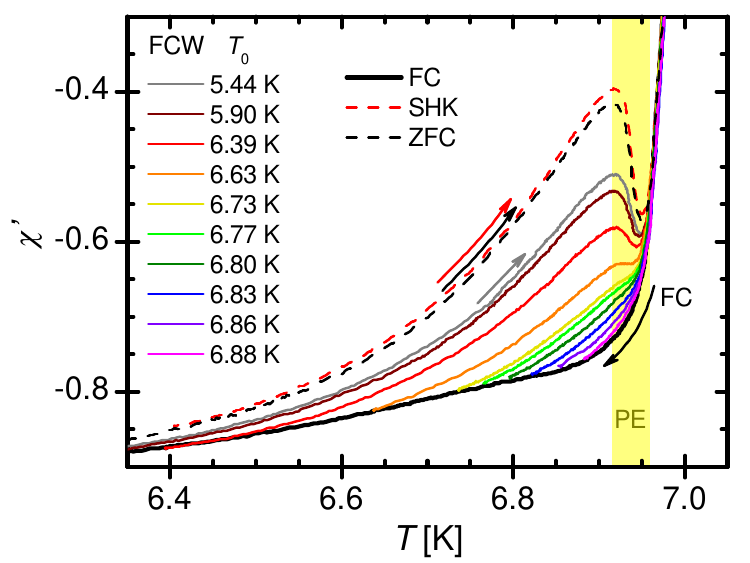}
\caption{Linear in-phase $ac$ susceptibility in a {\NS} single crystal at $H=1$~kOe, measured with $h_\mathrm{ac} = 10$~mOe, in FC (black) and FCW procedures [differing by their minimal temperature $T_{0}$]. Shaking the vortex state at low temperature with 1000 shaking cycles ($16$~Oe, $1$~kHz) installs a state which responds (dashed, red) similar to the ZFC experiment (dashed, black).}
\label{fig:exp}
\end{figure}

We first review the experimental signatures of thermal and dynamic history effects in {\NS} with a particular focus on the results reported in Ref.\ \sbonlinecite{Pasquini2008, Perez2011, Marziali2015, Marziali2017}.
Hysteresis in the response of $ac$-driven vortex systems have been observed in many superconductors, including traditional BCS materials \cite{Henderson1998, Xiao2004, Li2006}, high $T_c$ cuprates \cite{Valenzuela2000, Stamopoulos2002}, and other non-conventional superconductors \cite{Willa2015a}. 
In all those cases, a dependence of pinning on thermal and/or dynamical history has been reported and associated with different underlying vortex lattice configurations.

In the experiments reported in Refs.\ \sbonlinecite{Marziali2015, Marziali2017}, samples with thickness around $0.2$ mm and areas of several square millimeters, are exposed to a static field $H$ of the order of 1-5 kOe along the crystallographic $c$-axis (either after zero-field cooling, of by field-cooling through the superconducting transition $T_{c}\sim 7.2$ K). The response of the realized vortex configuration is then probed with a perturbative technique,  whereby the system is subjected to a small $ac$ field $h_\mathrm{ac} \!<\! 10~\mathrm{mOe}$ (also parallel to $H$) and its response is probed through the $ac$ susceptibility $\chi = \chi' + i \chi''$. The frequency-independent in-phase component $\chi'$ and the vanishing out-of-phase component $\chi''$ certify that the system is probed in the non-invasive Campbell regime \cite{Marziali2015sup}. Besides investigating different thermal histories of the vortex state upon thermal cycling, dynamic history effects are probed by shaking the system at specific temperatures $\Tsh$ with a strong $ac$ field of several Oersted (parallel to $H$). 

As shown in Fig. \ref{fig:exp}, $ac$ susceptibility measurements feature thermal hysteresis and dynamic history effects. 
The strongest diamagnetic response is observed when cooling the sample in the field (field cooling, FC; solid black line). 
Upon reaching a minimum temperature $T_{0}$ and subsequent warming (FCW), the response at any temperature $T > T_0$ deviates from the FC value and becomes less diamagnetic.
Distinct field-cooled warming curves are obtained along independent FC-FCW cycles (differing by $T_0$, colored solid lines) and the difference between FC and FCW curves grows with decreasing $T_{0}$. 
Moreover, for low enough $T_{0}$, FCW curves exhibit the peak effect (a strong reduction of the diamagnetic signal near $T_c$), not present in the FC curve.
For a zero-field-cooled (ZFC) state, the response is even less diamagnetic than in any of the FCW cases and displays a stronger peak effect (dashed black line). A similar response is observed after shaking the system with a large ac field (several Oe) at $\Tsh = 6.4$ K, below the peak effect (SHK, dashed red curve).

The prevalent picture for interpreting these findings is based on dislocations of the flux-line lattice: Following this argument, the vortex state upon field-cooling is characterized by a significant density of dislocations, hence improving the shielding of the state to an $ac$ perturbation. In contrast, the zero-field cooled state and the vortex state obtained after shaking feature a rather clean flux-line lattice, resulting in a less diamagnetic $ac$ response. Support for this interpretation come from small angle neutron scattering experiments\cite{Marziali2015, Marziali2017} resolving the mean spatial vortex lattice correlation length\cite{Yaron1995}.
Combined SANS and in-situ $ac$ susceptibility measurements in {\NS} single crystals showed that the FC configuration remains partially disordered down to low temperatures, as indicated by a short vortex lattice correlation length and associated with the presence of vortex lattice dislocations. After shaking, the system is driven into an ordered phase, free of dislocations and characterized by narrow, resolution-limited Bragg peaks. A similar ordering effect has been seen in STM experiments\cite{ChandraGanguli2015}. What remains unresolved in this dislocation-based picture is the observation that the changes in the susceptibility due to thermal cycling, do not result in a measurable change of the vortex lattice correlation lengths. 

\subsection{$Ac$ response in {\NS} revisited}\label{sec:revisted}
We propose a new interpretation of the experimental results on {\NS} from the strong-pinning perspective introduced in section \ref{sec:theory}. 
We shall see that the description of the linear Campbell and non-linear Bean dynamics within this framework provide ample support for an interpretation within the microscopic picture of strong vortex pinning.

First, the decreasing trend of the Campbell length upon cooling (below the peak-effect region), and the fact that $\lC{}_{,\mathrm{FCW}}>\lC{}_{,\mathrm{FC}}$ are indicative\cite{Willa2016} of the realization of Eq.\ \eqref{eq:casebprime} rather than Eq.\ \eqref{eq:caseb} when field-cooling the system, and suggest that $x_{+}$ shrinks with decreasing temperature. In this situation, the strong-pinning curvature \eqref{eq:alpha-sp} becomes maximal and, so does the diamagnetic response. When identifying $x_{+}(T)\approx f_{p}/\bar{C}$, the decrease in $x_{+}(T)$ can be attributed to an elastic stiffness $\bar{C}$ of the vortex lattice growing faster than the pinning force $f_{p}$ during cooling, a behavior consistent with an order-disorder transition upon warming.

Second, the decreasing pinning length $x_{+}$ upon cooling naturally leads to a new microscopic arrangement of vortices on the warming curve. Indeed, the force jump on the warming curve occurs at $\pm x_{+}(T_{0})$, which generically increases the Campbell length upon warming. The FCW response depends on $T_{0}$, and, as $T_{0}$ is lowered, it moves closer to the response obtained after shaking.

Furthermore, the application of a large shaking $ac$ field at low temperatures has two important consequences: On the one hand, after shaking the system, the vortex state is probed in a history-independent state, that is indistinguishable from the Bean critical state. And indeed, the Campbell response follows a universal curve upon warming; consistent with the susceptibility  measurements.
On the other hand, the shaking might be associated with a dynamic cleaning of dislocations. The increase of the vortex lattice correlation length, as revealed by SANS observations\cite{Marziali2015,Marziali2017}, indeed suggests that this effect is present here. One possible way to account for this phenomenon within the framework of strong vortex pinning is to evaluate a renormalized elasticity $\bar{C}$ due to vortex-lattice dislocations. Yet, this topic requires further investigations before drawing definite conclusions.

All these items support the interpretation of the experimental findings within the strong-pinning framework. Beyond this qualitative analysis, the hysteresis in Campbell response can be used to draw more quantitative conclusions and extract microscopic pinning parameters.

\section{Pinscape Spectroscopy}\label{sec:extraction}
In this section, we illustrate how the temperature evolution of pinning parameters can be extracted from a systematic study of the thermal hysteresis of the Campbell length. Specifically, the hysteresis of FC and FCW at the turning temperature $T_{0}$ allows to obtain the logarithmic derivatives $d \ln(f_{p})/dT |_{T_{0}}$ and $d \ln(x_{+})/dT |_{T_{0}}$, and hence provides direct access to the microscopic parameters $f_{p}$ and $x_{+}$. Below, we apply this procedure to experimental data shown Fig.\ \ref{fig:exp}, and reported in Ref.\ \sbonlinecite{Marziali2017}.

For simplicity, we assume that the pinned branch is almost linear $f_{\mathrm{pin}}(x)\approx (x/x_{+})f_{p}$ [with $f_{p}$ the defect's maximal (pin-breaking) force] and the force contribution from the unpinned branch may be neglected. This assumption is strictly valid only under very strong-pinning conditions, i.e., $\kappa \gg 1$, yet it yields an estimate for the more general case. Under this assumption, the above force term $\Delta f_{\mathrm{pin}}$ takes the simple form $\Delta f_{\mathrm{pin}%
}\approx \lbrack (x_{\mathrm{jp},+}+x_{\mathrm{jp},-})/x_{+}]f_{p}$. Combining this approximation with Eqs.\ \eqref{eq:Campbell-length} and \eqref{eq:alpha-sp} we find 
\begin{align}\label{eq:simplified-lC}
\lambda _{\mathrm{\scriptscriptstyle C}}^{-2}=\frac{2\pi n_{p}}{B\Phi _{0}}\frac{r_{\perp }(x_{\mathrm{jp},+}+x_{\mathrm{jp},-})f_{p}}{x_{+}}.
\end{align}
\begin{figure}[t]
\centering
\includegraphics[width = .47\textwidth]{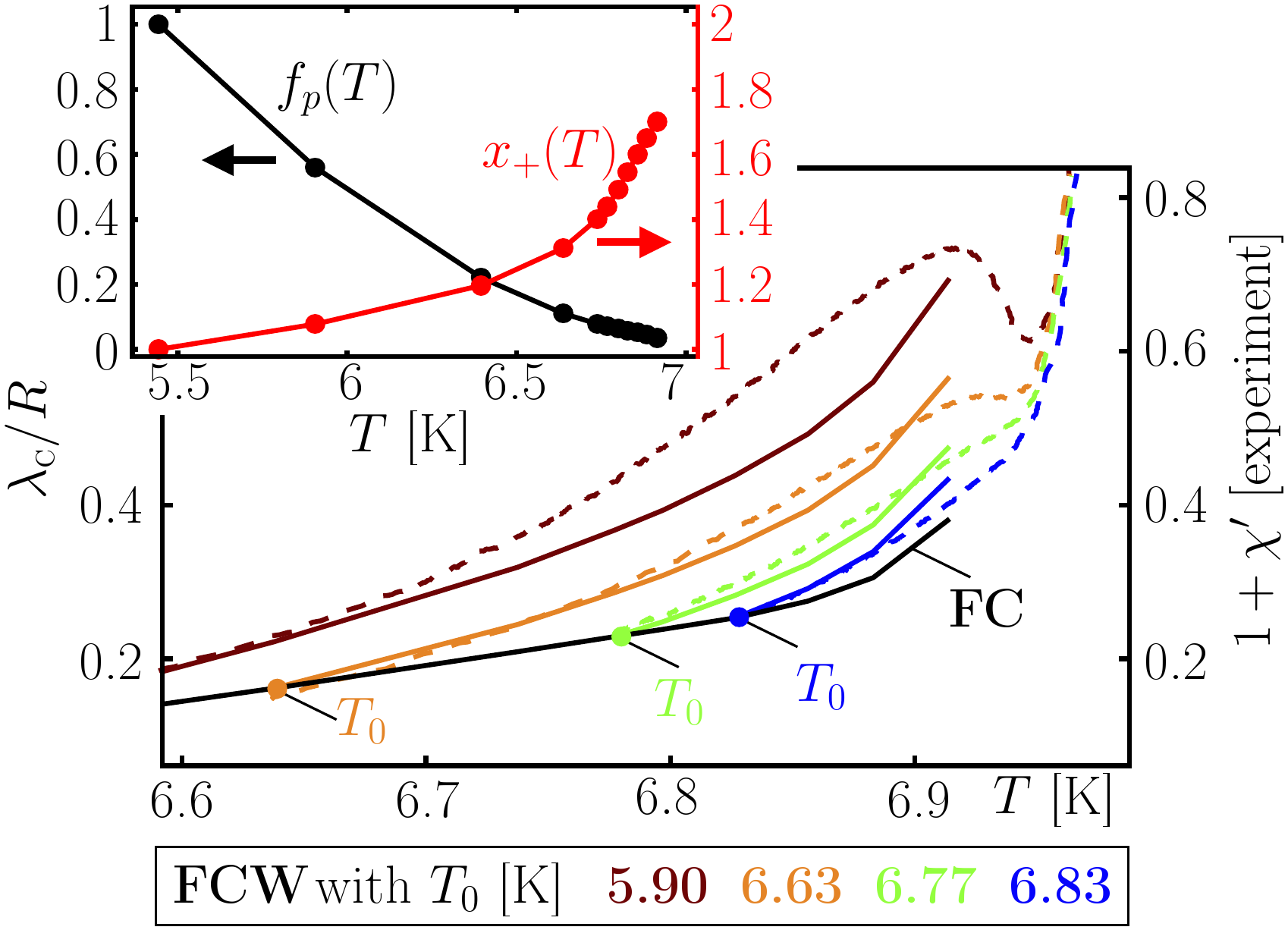}
\caption{Temperature dependence of the pin-breaking force $f_{p}$ and
the critical pinning length $x_{+}$ (inset) from the thermal cycling measurements in {\NS}. 
By using the relation $\lC \propto (1+\chi')$ and following Eqs.\ \eqref{eq:lC-c} and \eqref{eq:lC-w} with the extracted parameters, the $ac$ susceptibility (main figure, dashed, right axis) can be reconstructed (solid curve, left axis).}
\label{fig:extracted-pin-params}
\end{figure}
Let us discuss the situation of monotonically decreasing $x_{+}(T)$ (upon cooling) within the scheme discussed in section \ref{sec:theory} assuming the simplified form Eq.\ \eqref{eq:simplified-lC} for the Campbell length. On the cooling branch, we have $x_{\mathrm{jp},+}(T) = x_{\mathrm{jp},-}(T) = r_{\perp}(T) = x_{+}(T)$ (case \emph{b}). Inserting this dependence into Eq.\ \eqref{eq:simplified-lC}, we find
 \begin{align}  \label{eq:lC-c} \LC^{-2} \equiv \lC^{-2} \Big(\frac{16\pi n_{p}}{B \Phi_{0}}\Big)^{-1} = x_{+}(T) f_{p}(T),
\end{align}
where $\LC$ has been introduced for notational simplicity. Upon warming from a minimum temperature $T_{0}$, the jump position remains constant $x_{\mathrm{jp},\pm}(T) \!=\! r_{\perp}(T) \!=\! x_{+}(T_{0})$ (case \emph{a}), resulting in 
\begin{align}\label{eq:lC-w}
\LC^{-2} = \frac{x_{+}^{2}(T_{0})}{x_{+}(T)} f_{p}(T).
\end{align}
Away from $T_{0}$, above a certain temperature $T_{1}$ [defined through $x_{-}(T_{1}) = x_{+}(T_{0})$], the positions of the force jumps $x_{\mathrm{jp},\pm}(T) \!=\! r_{\perp}(T) \!=\! x_{-}(T)$ will acquire a temperature dependence again (case \emph{b}), leading to 
\begin{align}
   \LC^{-2} = \frac{x_{-}^{2}(T)}{x_{+}(T)}f_{p}(T).
\end{align}
The discussion in Sec.\ \ref{sec:revisted} suggests that $T_{1}$ is within the peak effect region, where the interpretation of the data is more challenging. However, the behavior near $T_{0}$ is very informative. Indeed upon taking a logarithmic derivatives of Eqs.\ \eqref{eq:lC-c} and  \eqref{eq:lC-w} we find 
\begin{align}\label{eq:term-c}
\left.\frac{d \ln (\LC^{-2})}{dT} \right|_{\mathrm{FC},T_{0}\;} &= \frac{d \ln (f_{p})}{dT} + \frac{d \ln (x_{+})}{dT}, \text{ and} \\
\label{eq:term-w}
\left.\frac{d \ln (\LC^{-2})}{dT} \right|_{\mathrm{FCW},T_{0}} &= \frac{d \ln (f_{p})}{dT} - \frac{d \ln (x_{+})}{dT},
\end{align}
for the cooling and warming branch, respectively. The inequality $d \ln (\LC^{-2})/dT|_{\mathrm{FC},T_{0}} \!>\! d \ln (\LC^{-2})/ dT|_{\mathrm{FCW},T_{0}}$ 
is another indicator that $x_{+}(T)$ increases with increasing temperature. 
Sum and difference of the two terms \eqref{eq:term-c} and \eqref{eq:term-w} grant direct access to the microscopic pinning parameters via $d \ln(f_{p})/dT$ and $d \ln(x_{+}) / dT$, respectively. Repeated hysteresis loops, taken at different $T_{0}$ therefore allows extracting the \emph{relative} temperature dependence of the pinning force $f_{p}$ and the pinning length $x_{+}$. 
Note that the \emph{absolute} value of the pinning parameters cannot be determined from this experiment, requiring complementary approaches, e.g. local probe experiments\cite{Embon2015}.

The abundance of thermal sweep data invites us to follow this route and extract the pinning parameters, and the result of this pinscape spectroscopy is summarized in Fig.\ \ref{fig:extracted-pin-params}. Each temperature marked by a point in the figure's inset derives from one thermal cycling experiment with the corresponding minimal temperature $T = T_{0}$. Because of the logarithmic nature of the expressions \eqref{eq:term-c} and \eqref{eq:term-w}, the relative changes are extracted from the data, i.e. $x_{+}$ and $f_{p}$ are normalized to the value extracted at the lowest $T_{0}$ (here $5.44~\mathrm{K}$). Figure \ref{fig:extracted-pin-params} overlays the Campbell response reconstructed from the pinning parameters with the measured $ac$ susceptibility data. The agreement extends well beyond the turning temperatures $T_{0}$ from which the pinning parameters are evaluated.

Having gained detailed information about the pinscape parameters from the data in the vicinity of the turning temperatures $T_{0}$, the linear $ac$ susceptibility can be reconstructed away from these temperatures, see the main panel of Fig.\ \ref{fig:extracted-pin-params}, and agrees well with the experimental data.

\section{Conclusion}

Within the strong-pinning paradigm, vortices occupy a multivalued pinning force profile. As this occupation is sensitive to the state preparation it may consequently evolve upon applying thermal cycles or large $ac$ magnetic field oscillations to the system. As a result, the macroscopic response of the pinned vortex state is known to show hysteretic behavior in macroscopic observables, e.g., in the linear $ac$ susceptibility (Campbell response). Here, we have derived a set of rules governing the microscopic occupation of the pinning force branches and its thermal evolution. Studying the implications of these rules on the Campbell response, we have laid out a path to evaluate the temperature evolution of the characteristic parameters defining the pinning problem at the microscopic scale.

We have shown that the history effects observed by $ac$ susceptibility measurements in {\NS} single crystals below the peak effect find a natural explanation within the Campbell response derived from strong-pinning theory. Most prominently, the hysteresis upon thermal cycling a field-cooled vortex state---up to now unexplained---naturally emerges from different occupations of the microscopic (multivalued) pinning force. For the same reason, a rather large difference between the field-cooled and zero-field cooled states is observed. Furthermore, the application of a large-amplitude shaking field pushes the system into a new state which is (a) independent of the state prior to shaking and (b) indistinguishable from a zero-field cooled state; two observations confirming the model's prediction.

With these evidences at hand, we have performed a quantitative analysis of the hysteresis between the field-cooled and field-cooled warming curves at different turning temperatures $T_{0}$ and have tested the new spectroscopic technique, whereby the temperature dependence of the pinning force and the pinning length were extracted from the $ac$ susceptibility data. While this differential technique only allows to probe relative changes, absolute values of the pinning parameters can be obtained by comparison with complementary spectroscopic approaches. In such a combination, our work opens the way for further systematic characterization of pinning landscapes in other vortex systems presenting hysteretic responses and compatible with the strong-pinning formalism.

\begin{acknowledgments}
The authors thank Vadim B.\ Geshkenbein, Gianni Blatter, and Victoria Bekeris for valuable discussions. This work was supported by the U.S. Department of Energy, Office of Science, Materials Sciences and Engineering Division,  Universidad de Buenos Aires and CONICET.\ R.W. acknowledges funding support from the Postdoc.Mobility fellowship of the SNSF.
\end{acknowledgments}

\vfill
\pagebreak
\appendix

\section{Turn-off process of the shaking field}\label{app:ramping}

\begin{figure}[tb]
\centering
\includegraphics[width = .46\textwidth]{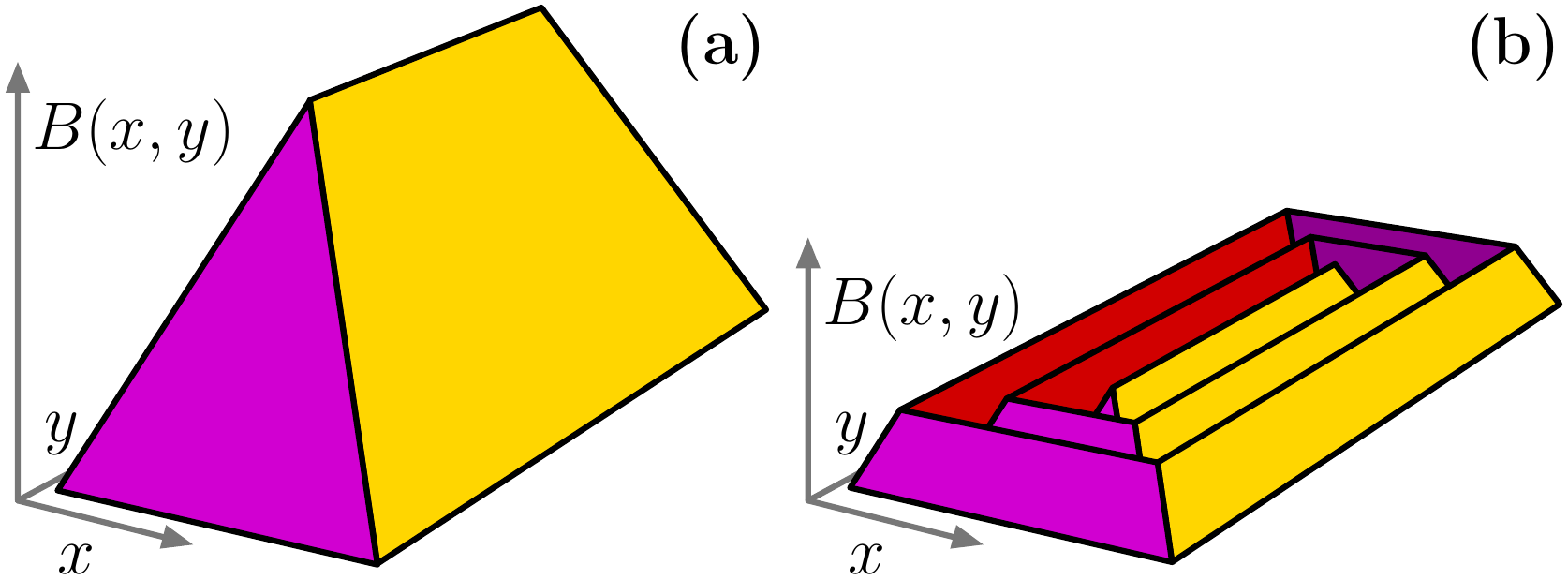}
\caption{A strong shaking field $h_\mathrm{ac}e^{-i \omega t}$ periodically induces a Bean critical state (hip-roof shaped vortex density). Panel (a) illustrates the field profile after turning the shaking field off abruptly, i.e. on a time scale faster than $1/\omega$. The system gets frozen in a Bean critical state. For a slow turn-off, panel (b), the system ends in a more complex state of local criticality.}
\label{fig:hip-roof}
\end{figure}

When turning the shaking field off abruptly $(\partial h_\mathrm{ac} / \partial t) \tau \gtrsim h_\mathrm{ac}$ (with $\tau = \omega^{-1}$ the period of the $ac$ field), the vortex system remains locked in a Bean critical state (single Bean triangle, rectangular hip roof) defined through critical currents, see Fig.\ \ref{fig:hip-roof}.
If the shaking field is turned off slower, the Bean penetration length changes only little within one oscillation cycle, i.e., $\delta \lambda_{\mathrm{\scriptscriptstyle B}} / \lambda_{\mathrm{\scriptscriptstyle B}} = (\tau / h_\mathrm{ac})(\partial h_\mathrm{ac} / \partial t) < 1$. For $\lC \ll \delta \lambda_{\mathrm{\scriptscriptstyle B}} \ll \lambda_{\mathrm{\scriptscriptstyle B}}$, the final vortex state corresponds to zig-zag shaped density with successive critical states defined through $\pm j_{c}$. The width of each Bean triangle is $\delta \lambda_{\mathrm{\scriptscriptstyle B}}$. After the shaking field is turned off, the Campbell response probes the critical state of the 'last' Bean triangle near the sample surface, since $\lC \!\ll\!  \delta \lambda_{\mathrm{\scriptscriptstyle B}}$ and hence the second triangle is deep inside the superconductor.
For even slower turn-off, i.e., $\delta \lambda_{\mathrm{\scriptscriptstyle B}} \ll \lC$, the vortex state relaxes to a force-free configuration where the force jump position $x_{\mathrm{jp},\pm} = (x_{+} + x_{-})/2$ generically differs from that of the field-cooled state $x_{\mathrm{jp},0}$.

\vfill
\pagebreak
\bibliographystyle{apsrev4-1-titles}
\bibliography{bulk-pinscape-spectroscopy}

\begin{thebibliography}{47}%
\makeatletter
\providecommand \@ifxundefined [1]{%
 \@ifx{#1\undefined}
}%
\providecommand \@ifnum [1]{%
 \ifnum #1\expandafter \@firstoftwo
 \else \expandafter \@secondoftwo
 \fi
}%
\providecommand \@ifx [1]{%
 \ifx #1\expandafter \@firstoftwo
 \else \expandafter \@secondoftwo
 \fi
}%
\providecommand \natexlab [1]{#1}%
\providecommand \enquote  [1]{``#1''}%
\providecommand \bibnamefont  [1]{#1}%
\providecommand \bibfnamefont [1]{#1}%
\providecommand \citenamefont [1]{#1}%
\providecommand \href@noop [0]{\@secondoftwo}%
\providecommand \href [0]{\begingroup \@sanitize@url \@href}%
\providecommand \@href[1]{\@@startlink{#1}\@@href}%
\providecommand \@@href[1]{\endgroup#1\@@endlink}%
\providecommand \@sanitize@url [0]{\catcode `\\12\catcode `\$12\catcode
  `\&12\catcode `\#12\catcode `\^12\catcode `\_12\catcode `\%12\relax}%
\providecommand \@@startlink[1]{}%
\providecommand \@@endlink[0]{}%
\providecommand \url  [0]{\begingroup\@sanitize@url \@url }%
\providecommand \@url [1]{\endgroup\@href {#1}{\urlprefix }}%
\providecommand \urlprefix  [0]{URL }%
\providecommand \Eprint [0]{\href }%
\providecommand \doibase [0]{http://doi.org/}%
\providecommand \selectlanguage [0]{\@gobble}%
\providecommand \bibinfo  [0]{\@secondoftwo}%
\providecommand \bibfield  [0]{\@secondoftwo}%
\providecommand \translation [1]{[#1]}%
\providecommand \BibitemOpen [0]{}%
\providecommand \bibitemStop [0]{}%
\providecommand \bibitemNoStop [0]{.\EOS\space}%
\providecommand \EOS [0]{\spacefactor3000\relax}%
\providecommand \BibitemShut  [1]{\csname bibitem#1\endcsname}%
\let\auto@bib@innerbib\@empty
\bibitem [{\citenamefont {Blatter}\ \emph {et~al.}(1994)\citenamefont
  {Blatter}, \citenamefont {Feigel'man}, \citenamefont {Geshkenbein},
  \citenamefont {Larkin},\ and\ \citenamefont {Vinokur}}]{Blatter1994}%
  \BibitemOpen
  \bibfield  {author} {\bibinfo {author} {\bibfnamefont {G.}~\bibnamefont
  {Blatter}}, \bibinfo {author} {\bibfnamefont {M.~V.}\ \bibnamefont
  {Feigel'man}}, \bibinfo {author} {\bibfnamefont {V.~B.}\ \bibnamefont
  {Geshkenbein}}, \bibinfo {author} {\bibfnamefont {A.~I.}\ \bibnamefont
  {Larkin}}, \ and\ \bibinfo {author} {\bibfnamefont {V.~M.}\ \bibnamefont
  {Vinokur}},\ }\bibfield  {title} {\emph {\bibinfo {title} {{Vortices in
  high-temperature superconductors}}},\ }\href@noop {} {\bibfield  {journal}
  {\bibinfo  {journal} {Review of Modern Physics}\ }\textbf {\bibinfo {volume}
  {66}},\ \bibinfo {pages} {1125} (\bibinfo {year} {1994})}\BibitemShut
  {NoStop}%
\bibitem [{\citenamefont {MacManus-Driscoll}\ \emph {et~al.}(2004)\citenamefont
  {MacManus-Driscoll}, \citenamefont {Foltyn}, \citenamefont {Jia},
  \citenamefont {Wang}, \citenamefont {Serquis}, \citenamefont {Maiorov},
  \citenamefont {Civale}, \citenamefont {Lin}, \citenamefont {Hawley},
  \citenamefont {Maley},\ and\ \citenamefont {Peterson}}]{MacManus2004}%
  \BibitemOpen
  \bibfield  {author} {\bibinfo {author} {\bibfnamefont {J.~L.}\ \bibnamefont
  {MacManus-Driscoll}}, \bibinfo {author} {\bibfnamefont {S.~R.}\ \bibnamefont
  {Foltyn}}, \bibinfo {author} {\bibfnamefont {Q.~X.}\ \bibnamefont {Jia}},
  \bibinfo {author} {\bibfnamefont {H.}~\bibnamefont {Wang}}, \bibinfo {author}
  {\bibfnamefont {A.}~\bibnamefont {Serquis}}, \bibinfo {author} {\bibfnamefont
  {B.}~\bibnamefont {Maiorov}}, \bibinfo {author} {\bibfnamefont
  {L.}~\bibnamefont {Civale}}, \bibinfo {author} {\bibfnamefont
  {Y.}~\bibnamefont {Lin}}, \bibinfo {author} {\bibfnamefont {M.~E.}\
  \bibnamefont {Hawley}}, \bibinfo {author} {\bibfnamefont {M.~P.}\
  \bibnamefont {Maley}}, \ and\ \bibinfo {author} {\bibfnamefont {D.~E.}\
  \bibnamefont {Peterson}},\ }\bibfield  {title} {\emph {\bibinfo {title}
  {Systematic enhancement of in-field critical current density with rare-earth
  ion size variance in superconducting rare-earth barium cuprate films}},\
  }\href {\doibase 10.1063/1.1766394} {\bibfield  {journal} {\bibinfo
  {journal} {Appl. Phys. Lett.}\ }\textbf {\bibinfo {volume} {84}},\ \bibinfo
  {pages} {5329} (\bibinfo {year} {2004})}\BibitemShut {NoStop}%
\bibitem [{\citenamefont {Haugan}\ \emph {et~al.}(2004)\citenamefont {Haugan},
  \citenamefont {Barnes}, \citenamefont {Wheeler}, \citenamefont
  {Meisenkothen},\ and\ \citenamefont {Sumption}}]{Haugan2004}%
  \BibitemOpen
  \bibfield  {author} {\bibinfo {author} {\bibfnamefont {T.}~\bibnamefont
  {Haugan}}, \bibinfo {author} {\bibfnamefont {P.~N.}\ \bibnamefont {Barnes}},
  \bibinfo {author} {\bibfnamefont {R.}~\bibnamefont {Wheeler}}, \bibinfo
  {author} {\bibfnamefont {F.}~\bibnamefont {Meisenkothen}}, \ and\ \bibinfo
  {author} {\bibfnamefont {M.}~\bibnamefont {Sumption}},\ }\bibfield  {title}
  {\emph {\bibinfo {title} {{Addition of nanoparticle dispersions to enhance
  flux pinning of the {$\mathrm{Y}\mathrm{Ba}_2\mathrm{Cu}_3\mathrm{O}_{7-x}$}
  superconductor}}},\ }\href {\doibase 10.1038/nature02792} {\bibfield
  {journal} {\bibinfo  {journal} {Nature}\ }\textbf {\bibinfo {volume} {430}},\
  \bibinfo {pages} {867} (\bibinfo {year} {2004})}\BibitemShut {NoStop}%
\bibitem [{\citenamefont {Kang}\ \emph {et~al.}(2006)\citenamefont {Kang},
  \citenamefont {Goyal}, \citenamefont {Li}, \citenamefont {Gapud},
  \citenamefont {Martin}, \citenamefont {Heatherly}, \citenamefont {Thompson},
  \citenamefont {Christen}, \citenamefont {List}, \citenamefont {Paranthaman},\
  and\ \citenamefont {Lee}}]{Kang2006}%
  \BibitemOpen
  \bibfield  {author} {\bibinfo {author} {\bibfnamefont {S.}~\bibnamefont
  {Kang}}, \bibinfo {author} {\bibfnamefont {A.}~\bibnamefont {Goyal}},
  \bibinfo {author} {\bibfnamefont {J.}~\bibnamefont {Li}}, \bibinfo {author}
  {\bibfnamefont {A.~A.}\ \bibnamefont {Gapud}}, \bibinfo {author}
  {\bibfnamefont {P.~M.}\ \bibnamefont {Martin}}, \bibinfo {author}
  {\bibfnamefont {L.}~\bibnamefont {Heatherly}}, \bibinfo {author}
  {\bibfnamefont {J.~R.}\ \bibnamefont {Thompson}}, \bibinfo {author}
  {\bibfnamefont {D.~K.}\ \bibnamefont {Christen}}, \bibinfo {author}
  {\bibfnamefont {F.~A.}\ \bibnamefont {List}}, \bibinfo {author}
  {\bibfnamefont {M.}~\bibnamefont {Paranthaman}}, \ and\ \bibinfo {author}
  {\bibfnamefont {D.~F.}\ \bibnamefont {Lee}},\ }\bibfield  {title} {\emph
  {\bibinfo {title} {{High-performance high-{$T_c$} superconducting wires}}},\
  }\href {\doibase 10.1126/science.1124872} {\bibfield  {journal} {\bibinfo
  {journal} {Science}\ }\textbf {\bibinfo {volume} {311}},\ \bibinfo {pages}
  {1911} (\bibinfo {year} {2006})}\BibitemShut {NoStop}%
\bibitem [{\citenamefont {Silhanek}\ \emph {et~al.}(2007)\citenamefont
  {Silhanek}, \citenamefont {Gillijns}, \citenamefont {Milo\ifmmode
  \check{s}\else \v{s}\fi{}evi\ifmmode~\acute{c}\else \'{c}\fi{}},
  \citenamefont {Volodin}, \citenamefont {Moshchalkov},\ and\ \citenamefont
  {Peeters}}]{Silhanek2007}%
  \BibitemOpen
  \bibfield  {author} {\bibinfo {author} {\bibfnamefont {A.~V.}\ \bibnamefont
  {Silhanek}}, \bibinfo {author} {\bibfnamefont {W.}~\bibnamefont {Gillijns}},
  \bibinfo {author} {\bibfnamefont {M.~V.}\ \bibnamefont {Milo\ifmmode
  \check{s}\else \v{s}\fi{}evi\ifmmode~\acute{c}\else \'{c}\fi{}}}, \bibinfo
  {author} {\bibfnamefont {A.}~\bibnamefont {Volodin}}, \bibinfo {author}
  {\bibfnamefont {V.~V.}\ \bibnamefont {Moshchalkov}}, \ and\ \bibinfo {author}
  {\bibfnamefont {F.~M.}\ \bibnamefont {Peeters}},\ }\bibfield  {title} {\emph
  {\bibinfo {title} {{Optimization of superconducting critical parameters by
  tuning the size and magnetization of arrays of magnetic dots}}},\ }\href
  {\doibase 10.1103/PhysRevB.76.100502} {\bibfield  {journal} {\bibinfo
  {journal} {Physical Review B}\ }\textbf {\bibinfo {volume} {76}},\ \bibinfo
  {pages} {100502} (\bibinfo {year} {2007})}\BibitemShut {NoStop}%
\bibitem [{\citenamefont {Gutierrez}\ \emph {et~al.}(2007)\citenamefont
  {Gutierrez}, \citenamefont {Llordes}, \citenamefont {Gazquez}, \citenamefont
  {Gibert}, \citenamefont {Roma}, \citenamefont {Pomar}, \citenamefont
  {Sandiumenge}, \citenamefont {Mestres}, \citenamefont {Puig},\ and\
  \citenamefont {Obradors}}]{Gutierrez2007}%
  \BibitemOpen
  \bibfield  {author} {\bibinfo {author} {\bibfnamefont {J.}~\bibnamefont
  {Gutierrez}}, \bibinfo {author} {\bibfnamefont {A.}~\bibnamefont {Llordes}},
  \bibinfo {author} {\bibfnamefont {J.}~\bibnamefont {Gazquez}}, \bibinfo
  {author} {\bibfnamefont {M.}~\bibnamefont {Gibert}}, \bibinfo {author}
  {\bibfnamefont {N.}~\bibnamefont {Roma}}, \bibinfo {author} {\bibfnamefont
  {A.}~\bibnamefont {Pomar}}, \bibinfo {author} {\bibfnamefont
  {F.}~\bibnamefont {Sandiumenge}}, \bibinfo {author} {\bibfnamefont
  {N.}~\bibnamefont {Mestres}}, \bibinfo {author} {\bibfnamefont
  {T.}~\bibnamefont {Puig}}, \ and\ \bibinfo {author} {\bibfnamefont
  {X.}~\bibnamefont {Obradors}},\ }\bibfield  {title} {\emph {\bibinfo {title}
  {{Strong isotropic flux pinning in solution-derived
  {$\mathrm{Y}\mathrm{Ba}_2\mathrm{Cu}_3\mathrm{O}_{7-x}$} nanocomposite
  superconductor films}}},\ }\href {\doibase 10.1038/nmat1893} {\bibfield
  {journal} {\bibinfo  {journal} {Nature Materials}\ }\textbf {\bibinfo
  {volume} {6}},\ \bibinfo {pages} {367} (\bibinfo {year} {2007})}\BibitemShut
  {NoStop}%
\bibitem [{\citenamefont {Maiorov}\ \emph {et~al.}(2009)\citenamefont
  {Maiorov}, \citenamefont {Baily}, \citenamefont {Zhou}, \citenamefont
  {Ugurlu}, \citenamefont {Kennison}, \citenamefont {Dowden}, \citenamefont
  {Holesinger}, \citenamefont {Foltyn},\ and\ \citenamefont
  {Civale}}]{Maiorov2009}%
  \BibitemOpen
  \bibfield  {author} {\bibinfo {author} {\bibfnamefont {B.}~\bibnamefont
  {Maiorov}}, \bibinfo {author} {\bibfnamefont {S.~A.}\ \bibnamefont {Baily}},
  \bibinfo {author} {\bibfnamefont {H.}~\bibnamefont {Zhou}}, \bibinfo {author}
  {\bibfnamefont {O.}~\bibnamefont {Ugurlu}}, \bibinfo {author} {\bibfnamefont
  {J.~A.}\ \bibnamefont {Kennison}}, \bibinfo {author} {\bibfnamefont {P.~C.}\
  \bibnamefont {Dowden}}, \bibinfo {author} {\bibfnamefont {T.~G.}\
  \bibnamefont {Holesinger}}, \bibinfo {author} {\bibfnamefont {S.~R.}\
  \bibnamefont {Foltyn}}, \ and\ \bibinfo {author} {\bibfnamefont
  {L.}~\bibnamefont {Civale}},\ }\bibfield  {title} {\emph {\bibinfo {title}
  {{Synergetic combination of different types of defect to optimize pinning
  landscape using BaZrO3-doped YBa2Cu3O7}}},\ }\href
  {https://doi.org/10.1038/nmat2408} {\bibfield  {journal} {\bibinfo  {journal}
  {Nature Materials}\ }\textbf {\bibinfo {volume} {8}},\ \bibinfo {pages} {398
  EP} (\bibinfo {year} {2009})}\BibitemShut {NoStop}%
\bibitem [{\citenamefont {Polat}\ \emph {et~al.}(2011)\citenamefont {Polat},
  \citenamefont {Sinclair}, \citenamefont {Zuev}, \citenamefont {Thompson},
  \citenamefont {Christen}, \citenamefont {Cook}, \citenamefont {Kumar},
  \citenamefont {Chen},\ and\ \citenamefont {Selvamanickam}}]{Polat2011}%
  \BibitemOpen
  \bibfield  {author} {\bibinfo {author} {\bibfnamefont {O.}~\bibnamefont
  {Polat}}, \bibinfo {author} {\bibfnamefont {J.~W.}\ \bibnamefont {Sinclair}},
  \bibinfo {author} {\bibfnamefont {Y.~L.}\ \bibnamefont {Zuev}}, \bibinfo
  {author} {\bibfnamefont {J.~R.}\ \bibnamefont {Thompson}}, \bibinfo {author}
  {\bibfnamefont {D.~K.}\ \bibnamefont {Christen}}, \bibinfo {author}
  {\bibfnamefont {S.~W.}\ \bibnamefont {Cook}}, \bibinfo {author}
  {\bibfnamefont {D.}~\bibnamefont {Kumar}}, \bibinfo {author} {\bibfnamefont
  {Y.}~\bibnamefont {Chen}}, \ and\ \bibinfo {author} {\bibfnamefont
  {V.}~\bibnamefont {Selvamanickam}},\ }\bibfield  {title} {\emph {\bibinfo
  {title} {{Thickness dependence of magnetic relaxation and {$E$-$J$}
  characteristics in superconducting
  {($\mathrm{Gd}$-$\mathrm{Y}$)$\mathrm{Ba}\mathrm{Cu}$-$\mathrm{O}$} films
  with strong vortex pinning}}},\ }\href {\doibase 10.1103/PhysRevB.84.024519}
  {\bibfield  {journal} {\bibinfo  {journal} {Physical Review B}\ }\textbf
  {\bibinfo {volume} {84}},\ \bibinfo {pages} {024519} (\bibinfo {year}
  {2011})}\BibitemShut {NoStop}%
\bibitem [{\citenamefont {Miura}\ \emph {et~al.}(2013)\citenamefont {Miura},
  \citenamefont {Maiorov}, \citenamefont {Kato}, \citenamefont {Shimode},
  \citenamefont {Wada}, \citenamefont {Adachi},\ and\ \citenamefont
  {Tanabe}}]{Miura2013b}%
  \BibitemOpen
  \bibfield  {author} {\bibinfo {author} {\bibfnamefont {M.}~\bibnamefont
  {Miura}}, \bibinfo {author} {\bibfnamefont {B.}~\bibnamefont {Maiorov}},
  \bibinfo {author} {\bibfnamefont {T.}~\bibnamefont {Kato}}, \bibinfo {author}
  {\bibfnamefont {T.}~\bibnamefont {Shimode}}, \bibinfo {author} {\bibfnamefont
  {K.}~\bibnamefont {Wada}}, \bibinfo {author} {\bibfnamefont {S.}~\bibnamefont
  {Adachi}}, \ and\ \bibinfo {author} {\bibfnamefont {K.}~\bibnamefont
  {Tanabe}},\ }\bibfield  {title} {\emph {\bibinfo {title} {{Strongly enhanced
  flux pinning in one-step deposition of {BaFe$_2$(As$_{0.66}$P$_{0.33}$)$_2$}
  superconductor films with uniformly dispersed {BaZrO$_3$} nanoparticles}}},\
  }\href@noop {} {\bibfield  {journal} {\bibinfo  {journal} {Nature
  Communications}\ }\textbf {\bibinfo {volume} {4}},\ \bibinfo {pages} {2499}
  (\bibinfo {year} {2013})}\BibitemShut {NoStop}%
\bibitem [{\citenamefont {Ray}\ \emph {et~al.}(2013)\citenamefont {Ray},
  \citenamefont {Olson~Reichhardt}, \citenamefont {Jank\'o},\ and\
  \citenamefont {Reichhardt}}]{Ray2013}%
  \BibitemOpen
  \bibfield  {author} {\bibinfo {author} {\bibfnamefont {D.}~\bibnamefont
  {Ray}}, \bibinfo {author} {\bibfnamefont {C.~J.}\ \bibnamefont
  {Olson~Reichhardt}}, \bibinfo {author} {\bibfnamefont {B.}~\bibnamefont
  {Jank\'o}}, \ and\ \bibinfo {author} {\bibfnamefont {C.}~\bibnamefont
  {Reichhardt}},\ }\bibfield  {title} {\emph {\bibinfo {title} {{Strongly
  Enhanced Pinning of Magnetic Vortices in Type-II Superconductors by Conformal
  Crystal Arrays}}},\ }\href {\doibase 10.1103/PhysRevLett.110.267001}
  {\bibfield  {journal} {\bibinfo  {journal} {Physical Review Letters}\
  }\textbf {\bibinfo {volume} {110}},\ \bibinfo {pages} {267001} (\bibinfo
  {year} {2013})}\BibitemShut {NoStop}%
\bibitem [{\citenamefont {Kwok}\ \emph {et~al.}(2016)\citenamefont {Kwok},
  \citenamefont {Welp}, \citenamefont {Glatz}, \citenamefont {Koshelev},
  \citenamefont {Kihlstrom},\ and\ \citenamefont {Crabtree}}]{Kwok2016}%
  \BibitemOpen
  \bibfield  {author} {\bibinfo {author} {\bibfnamefont {W.-K.}\ \bibnamefont
  {Kwok}}, \bibinfo {author} {\bibfnamefont {U.}~\bibnamefont {Welp}}, \bibinfo
  {author} {\bibfnamefont {A.}~\bibnamefont {Glatz}}, \bibinfo {author}
  {\bibfnamefont {A.~E.}\ \bibnamefont {Koshelev}}, \bibinfo {author}
  {\bibfnamefont {K.~J.}\ \bibnamefont {Kihlstrom}}, \ and\ \bibinfo {author}
  {\bibfnamefont {G.~W.}\ \bibnamefont {Crabtree}},\ }\bibfield  {title} {\emph
  {\bibinfo {title} {{Vortices in high-performance high-temperature
  superconductors}}},\ }\href@noop {} {\bibfield  {journal} {\bibinfo
  {journal} {Reports on Progress in Physics}\ }\textbf {\bibinfo {volume}
  {79}},\ \bibinfo {pages} {116501} (\bibinfo {year} {2016})}\BibitemShut
  {NoStop}%
\bibitem [{\citenamefont {Sadovskyy}\ \emph {et~al.}(2016)\citenamefont
  {Sadovskyy}, \citenamefont {Jia}, \citenamefont {Leroux}, \citenamefont
  {Kwon}, \citenamefont {Hu}, \citenamefont {Fang}, \citenamefont {Chaparro},
  \citenamefont {Zhu}, \citenamefont {Welp}, \citenamefont {Zuo}, \citenamefont
  {Zhang}, \citenamefont {Nakasaki}, \citenamefont {Selvamanickam},
  \citenamefont {Crabtree}, \citenamefont {Koshelev}, \citenamefont {Glatz},\
  and\ \citenamefont {Kwok}}]{Sadovskyy2016b}%
  \BibitemOpen
  \bibfield  {author} {\bibinfo {author} {\bibfnamefont {I.~A.}\ \bibnamefont
  {Sadovskyy}}, \bibinfo {author} {\bibfnamefont {Y.}~\bibnamefont {Jia}},
  \bibinfo {author} {\bibfnamefont {M.}~\bibnamefont {Leroux}}, \bibinfo
  {author} {\bibfnamefont {J.}~\bibnamefont {Kwon}}, \bibinfo {author}
  {\bibfnamefont {H.}~\bibnamefont {Hu}}, \bibinfo {author} {\bibfnamefont
  {L.}~\bibnamefont {Fang}}, \bibinfo {author} {\bibfnamefont {C.}~\bibnamefont
  {Chaparro}}, \bibinfo {author} {\bibfnamefont {S.}~\bibnamefont {Zhu}},
  \bibinfo {author} {\bibfnamefont {U.}~\bibnamefont {Welp}}, \bibinfo {author}
  {\bibfnamefont {J.-M.}\ \bibnamefont {Zuo}}, \bibinfo {author} {\bibfnamefont
  {Y.}~\bibnamefont {Zhang}}, \bibinfo {author} {\bibfnamefont
  {R.}~\bibnamefont {Nakasaki}}, \bibinfo {author} {\bibfnamefont
  {V.}~\bibnamefont {Selvamanickam}}, \bibinfo {author} {\bibfnamefont {G.~W.}\
  \bibnamefont {Crabtree}}, \bibinfo {author} {\bibfnamefont {A.~E.}\
  \bibnamefont {Koshelev}}, \bibinfo {author} {\bibfnamefont {A.}~\bibnamefont
  {Glatz}}, \ and\ \bibinfo {author} {\bibfnamefont {W.-K.}\ \bibnamefont
  {Kwok}},\ }\bibfield  {title} {\emph {\bibinfo {title} {Toward
  superconducting critical current by design}},\ }\href@noop {} {\bibfield
  {journal} {\bibinfo  {journal} {Advanced Materials}\ }\textbf {\bibinfo
  {volume} {28}},\ \bibinfo {pages} {4593} (\bibinfo {year}
  {2016})}\BibitemShut {NoStop}%
\bibitem [{\citenamefont {Le~Thien}\ \emph {et~al.}(2017)\citenamefont
  {Le~Thien}, \citenamefont {McDermott}, \citenamefont {Reichhardt},\ and\
  \citenamefont {Reichhardt}}]{LeThien2017}%
  \BibitemOpen
  \bibfield  {author} {\bibinfo {author} {\bibfnamefont {Q.}~\bibnamefont
  {Le~Thien}}, \bibinfo {author} {\bibfnamefont {D.}~\bibnamefont {McDermott}},
  \bibinfo {author} {\bibfnamefont {C.~J.~O.}\ \bibnamefont {Reichhardt}}, \
  and\ \bibinfo {author} {\bibfnamefont {C.}~\bibnamefont {Reichhardt}},\
  }\bibfield  {title} {\emph {\bibinfo {title} {{Enhanced pinning for vortices
  in hyperuniform pinning arrays and emergent hyperuniform vortex
  configurations with quenched disorder}}},\ }\href {\doibase
  10.1103/PhysRevB.96.094516} {\bibfield  {journal} {\bibinfo  {journal}
  {Physical Review B}\ }\textbf {\bibinfo {volume} {96}},\ \bibinfo {pages}
  {094516} (\bibinfo {year} {2017})}\BibitemShut {NoStop}%
\bibitem [{\citenamefont {Eley}\ \emph {et~al.}(2018)\citenamefont {Eley},
  \citenamefont {Willa}, \citenamefont {Miura}, \citenamefont {Sato},
  \citenamefont {Leroux}, \citenamefont {Henry},\ and\ \citenamefont
  {Civale}}]{Eley2018a}%
  \BibitemOpen
  \bibfield  {author} {\bibinfo {author} {\bibfnamefont {S.}~\bibnamefont
  {Eley}}, \bibinfo {author} {\bibfnamefont {R.}~\bibnamefont {Willa}},
  \bibinfo {author} {\bibfnamefont {M.}~\bibnamefont {Miura}}, \bibinfo
  {author} {\bibfnamefont {M.}~\bibnamefont {Sato}}, \bibinfo {author}
  {\bibfnamefont {M.}~\bibnamefont {Leroux}}, \bibinfo {author} {\bibfnamefont
  {M.~D.}\ \bibnamefont {Henry}}, \ and\ \bibinfo {author} {\bibfnamefont
  {L.}~\bibnamefont {Civale}},\ }\bibfield  {title} {\emph {\bibinfo {title}
  {{Accelerated vortex dynamics across the magnetic 3D-to-2D crossover in
  disordered superconductors}}},\ }\href {\doibase 10.1038/s41535-018-0108-1}
  {\bibfield  {journal} {\bibinfo  {journal} {npj Quantum Materials}\ }\textbf
  {\bibinfo {volume} {3}},\ \bibinfo {pages} {37} (\bibinfo {year}
  {2018})}\BibitemShut {NoStop}%
\bibitem [{\citenamefont {Labusch}(1969)}]{Labusch1969}%
  \BibitemOpen
  \bibfield  {author} {\bibinfo {author} {\bibfnamefont {R.}~\bibnamefont
  {Labusch}},\ }\bibfield  {title} {\emph {\bibinfo {title} {{Calculation of
  the critical field gradient in type-II superconductors}}},\ }\href@noop {}
  {\bibfield  {journal} {\bibinfo  {journal} {Crystal Lattice Defects}\
  }\textbf {\bibinfo {volume} {1}},\ \bibinfo {pages} {1} (\bibinfo {year}
  {1969})}\BibitemShut {NoStop}%
\bibitem [{\citenamefont {Larkin}\ and\ \citenamefont
  {Ovchinnikov}(1979)}]{Larkin1979}%
  \BibitemOpen
  \bibfield  {author} {\bibinfo {author} {\bibfnamefont {A.~I.}\ \bibnamefont
  {Larkin}}\ and\ \bibinfo {author} {\bibfnamefont {Y.~N.}\ \bibnamefont
  {Ovchinnikov}},\ }\bibfield  {title} {\emph {\bibinfo {title} {{Pinning in
  type-II superconductors}}},\ }\href@noop {} {\bibfield  {journal} {\bibinfo
  {journal} {Journal of Low Temperature Physics}\ }\textbf {\bibinfo {volume}
  {34}},\ \bibinfo {pages} {409} (\bibinfo {year} {1979})}\BibitemShut
  {NoStop}%
\bibitem [{\citenamefont {Blatter}\ \emph {et~al.}(2004)\citenamefont
  {Blatter}, \citenamefont {Geshkenbein},\ and\ \citenamefont
  {Koopmann}}]{Blatter2004}%
  \BibitemOpen
  \bibfield  {author} {\bibinfo {author} {\bibfnamefont {G.}~\bibnamefont
  {Blatter}}, \bibinfo {author} {\bibfnamefont {V.~B.}\ \bibnamefont
  {Geshkenbein}}, \ and\ \bibinfo {author} {\bibfnamefont {J.~A.~G.}\
  \bibnamefont {Koopmann}},\ }\bibfield  {title} {\emph {\bibinfo {title}
  {{Weak to strong pinning crossover}}},\ }\href {\doibase
  10.1103/PhysRevLett.92.067009} {\bibfield  {journal} {\bibinfo  {journal}
  {Physical Review Letters}\ }\textbf {\bibinfo {volume} {92}},\ \bibinfo
  {pages} {067009} (\bibinfo {year} {2004})}\BibitemShut {NoStop}%
\bibitem [{\citenamefont {Thomann}\ \emph {et~al.}(2012)\citenamefont
  {Thomann}, \citenamefont {Geshkenbein},\ and\ \citenamefont
  {Blatter}}]{Thomann2012}%
  \BibitemOpen
  \bibfield  {author} {\bibinfo {author} {\bibfnamefont {A.~U.}\ \bibnamefont
  {Thomann}}, \bibinfo {author} {\bibfnamefont {V.~B.}\ \bibnamefont
  {Geshkenbein}}, \ and\ \bibinfo {author} {\bibfnamefont {G.}~\bibnamefont
  {Blatter}},\ }\bibfield  {title} {\emph {\bibinfo {title} {{Dynamical aspects
  of strong pinning of magnetic vortices in type-II superconductors}}},\ }\href
  {\doibase 10.1103/PhysRevLett.108.217001} {\bibfield  {journal} {\bibinfo
  {journal} {Physical Review Letters}\ }\textbf {\bibinfo {volume} {108}},\
  \bibinfo {pages} {217001} (\bibinfo {year} {2012})}\BibitemShut {NoStop}%
\bibitem [{\citenamefont {Thomann}\ \emph {et~al.}(2017)\citenamefont
  {Thomann}, \citenamefont {Geshkenbein},\ and\ \citenamefont
  {Blatter}}]{Thomann2017}%
  \BibitemOpen
  \bibfield  {author} {\bibinfo {author} {\bibfnamefont {A.~U.}\ \bibnamefont
  {Thomann}}, \bibinfo {author} {\bibfnamefont {V.~B.}\ \bibnamefont
  {Geshkenbein}}, \ and\ \bibinfo {author} {\bibfnamefont {G.}~\bibnamefont
  {Blatter}},\ }\bibfield  {title} {\emph {\bibinfo {title} {{Vortex dynamics
  in type-II superconductors under strong pinning conditions}}},\ }\href
  {\doibase 10.1103/PhysRevB.96.144516} {\bibfield  {journal} {\bibinfo
  {journal} {Physical Review B}\ }\textbf {\bibinfo {volume} {96}},\ \bibinfo
  {pages} {144516} (\bibinfo {year} {2017})}\BibitemShut {NoStop}%
\bibitem [{\citenamefont {Buchacek}\ \emph {et~al.}(2018)\citenamefont
  {Buchacek}, \citenamefont {Willa}, \citenamefont {Geshkenbein},\ and\
  \citenamefont {Blatter}}]{Buchacek2018a}%
  \BibitemOpen
  \bibfield  {author} {\bibinfo {author} {\bibfnamefont {M.}~\bibnamefont
  {Buchacek}}, \bibinfo {author} {\bibfnamefont {R.}~\bibnamefont {Willa}},
  \bibinfo {author} {\bibfnamefont {V.~B.}\ \bibnamefont {Geshkenbein}}, \ and\
  \bibinfo {author} {\bibfnamefont {G.}~\bibnamefont {Blatter}},\ }\bibfield
  {title} {\emph {\bibinfo {title} {{Persistence of pinning and creep beyond
  critical drive within the strong pinning paradigm}}},\ }\href {\doibase
  10.1103/PhysRevB.98.094510} {\bibfield  {journal} {\bibinfo  {journal}
  {Physical Review B}\ }\textbf {\bibinfo {volume} {98}},\ \bibinfo {pages}
  {094510} (\bibinfo {year} {2018})}\BibitemShut {NoStop}%
\bibitem [{\citenamefont {Willa}\ \emph
  {et~al.}(2015{\natexlab{a}})\citenamefont {Willa}, \citenamefont
  {Geshkenbein}, \citenamefont {Prozorov},\ and\ \citenamefont
  {Blatter}}]{Willa2015a}%
  \BibitemOpen
  \bibfield  {author} {\bibinfo {author} {\bibfnamefont {R.}~\bibnamefont
  {Willa}}, \bibinfo {author} {\bibfnamefont {V.~B.}\ \bibnamefont
  {Geshkenbein}}, \bibinfo {author} {\bibfnamefont {R.}~\bibnamefont
  {Prozorov}}, \ and\ \bibinfo {author} {\bibfnamefont {G.}~\bibnamefont
  {Blatter}},\ }\bibfield  {title} {\emph {\bibinfo {title} {{Campbell response
  in type-II superconductors under strong pinning conditions}}},\ }\href
  {\doibase 10.1103/PhysRevLett.115.207001} {\bibfield  {journal} {\bibinfo
  {journal} {Physical Review Letters}\ }\textbf {\bibinfo {volume} {115}},\
  \bibinfo {pages} {207001} (\bibinfo {year} {2015}{\natexlab{a}})}\BibitemShut
  {NoStop}%
\bibitem [{\citenamefont {Willa}\ \emph
  {et~al.}(2015{\natexlab{b}})\citenamefont {Willa}, \citenamefont
  {Geshkenbein},\ and\ \citenamefont {Blatter}}]{Willa2015b}%
  \BibitemOpen
  \bibfield  {author} {\bibinfo {author} {\bibfnamefont {R.}~\bibnamefont
  {Willa}}, \bibinfo {author} {\bibfnamefont {V.~B.}\ \bibnamefont
  {Geshkenbein}}, \ and\ \bibinfo {author} {\bibfnamefont {G.}~\bibnamefont
  {Blatter}},\ }\bibfield  {title} {\emph {\bibinfo {title} {{Campbell
  penetration in the critical state of type-II superconductors}}},\ }\href
  {\doibase 10.1103/PhysRevB.92.134501} {\bibfield  {journal} {\bibinfo
  {journal} {Physical Review B}\ }\textbf {\bibinfo {volume} {92}},\ \bibinfo
  {pages} {134501} (\bibinfo {year} {2015}{\natexlab{b}})}\BibitemShut
  {NoStop}%
\bibitem [{\citenamefont {Willa}\ \emph {et~al.}(2016)\citenamefont {Willa},
  \citenamefont {Geshkenbein},\ and\ \citenamefont {Blatter}}]{Willa2016}%
  \BibitemOpen
  \bibfield  {author} {\bibinfo {author} {\bibfnamefont {R.}~\bibnamefont
  {Willa}}, \bibinfo {author} {\bibfnamefont {V.~B.}\ \bibnamefont
  {Geshkenbein}}, \ and\ \bibinfo {author} {\bibfnamefont {G.}~\bibnamefont
  {Blatter}},\ }\bibfield  {title} {\emph {\bibinfo {title} {{Probing the
  pinning landscape in type-II superconductors via Campbell penetration
  depth}}},\ }\href {\doibase 10.1103/PhysRevB.93.064515} {\bibfield  {journal}
  {\bibinfo  {journal} {Physical Review B}\ }\textbf {\bibinfo {volume} {93}},\
  \bibinfo {pages} {064515} (\bibinfo {year} {2016})}\BibitemShut {NoStop}%
\bibitem [{\citenamefont {Pasquini}\ \emph {et~al.}(2008)\citenamefont
  {Pasquini}, \citenamefont {Daroca}, \citenamefont {Chiliotte}, \citenamefont
  {Lozano},\ and\ \citenamefont {Bekeris}}]{Pasquini2008}%
  \BibitemOpen
  \bibfield  {author} {\bibinfo {author} {\bibfnamefont {G.}~\bibnamefont
  {Pasquini}}, \bibinfo {author} {\bibfnamefont {D.~P.}\ \bibnamefont
  {Daroca}}, \bibinfo {author} {\bibfnamefont {C.}~\bibnamefont {Chiliotte}},
  \bibinfo {author} {\bibfnamefont {G.~S.}\ \bibnamefont {Lozano}}, \ and\
  \bibinfo {author} {\bibfnamefont {V.}~\bibnamefont {Bekeris}},\ }\bibfield
  {title} {\emph {\bibinfo {title} {{Ordered, disordered, and coexistent stable
  vortex lattices in ${\mathrm{NbSe}}_{2}$ single crystals}}},\ }\href
  {\doibase 10.1103/PhysRevLett.100.247003} {\bibfield  {journal} {\bibinfo
  {journal} {Physical Review Letters}\ }\textbf {\bibinfo {volume} {100}},\
  \bibinfo {pages} {247003} (\bibinfo {year} {2008})}\BibitemShut {NoStop}%
\bibitem [{\citenamefont {P\'erez~Daroca}\ \emph {et~al.}(2011)\citenamefont
  {P\'erez~Daroca}, \citenamefont {Pasquini}, \citenamefont {Lozano},\ and\
  \citenamefont {Bekeris}}]{Perez2011}%
  \BibitemOpen
  \bibfield  {author} {\bibinfo {author} {\bibfnamefont {D.}~\bibnamefont
  {P\'erez~Daroca}}, \bibinfo {author} {\bibfnamefont {G.}~\bibnamefont
  {Pasquini}}, \bibinfo {author} {\bibfnamefont {G.~S.}\ \bibnamefont
  {Lozano}}, \ and\ \bibinfo {author} {\bibfnamefont {V.}~\bibnamefont
  {Bekeris}},\ }\bibfield  {title} {\emph {\bibinfo {title} {{Dynamics of
  superconducting vortices driven by oscillatory forces in the plastic-flow
  regime}}},\ }\href {\doibase 10.1103/PhysRevB.84.012508} {\bibfield
  {journal} {\bibinfo  {journal} {Physical Review B}\ }\textbf {\bibinfo
  {volume} {84}},\ \bibinfo {pages} {012508} (\bibinfo {year}
  {2011})}\BibitemShut {NoStop}%
\bibitem [{\citenamefont {Marziali~Berm\'udez}\ \emph
  {et~al.}(2015)\citenamefont {Marziali~Berm\'udez}, \citenamefont {Eskildsen},
  \citenamefont {Bartkowiak}, \citenamefont {Nagy}, \citenamefont {Bekeris},\
  and\ \citenamefont {Pasquini}}]{Marziali2015}%
  \BibitemOpen
  \bibfield  {author} {\bibinfo {author} {\bibfnamefont {M.}~\bibnamefont
  {Marziali~Berm\'udez}}, \bibinfo {author} {\bibfnamefont {M.~R.}\
  \bibnamefont {Eskildsen}}, \bibinfo {author} {\bibfnamefont {M.}~\bibnamefont
  {Bartkowiak}}, \bibinfo {author} {\bibfnamefont {G.}~\bibnamefont {Nagy}},
  \bibinfo {author} {\bibfnamefont {V.}~\bibnamefont {Bekeris}}, \ and\
  \bibinfo {author} {\bibfnamefont {G.}~\bibnamefont {Pasquini}},\ }\bibfield
  {title} {\emph {\bibinfo {title} {Dynamic reorganization of vortex matter
  into partially disordered lattices}},\ }\href {\doibase
  10.1103/PhysRevLett.115.067001} {\bibfield  {journal} {\bibinfo  {journal}
  {Physical Review Letters}\ }\textbf {\bibinfo {volume} {115}},\ \bibinfo
  {pages} {067001} (\bibinfo {year} {2015})}\BibitemShut {NoStop}%
\bibitem [{\citenamefont {Marziali~Berm\'udez}\ \emph
  {et~al.}(2017)\citenamefont {Marziali~Berm\'udez}, \citenamefont {Louden},
  \citenamefont {Eskildsen}, \citenamefont {Dewhurst}, \citenamefont
  {Bekeris},\ and\ \citenamefont {Pasquini}}]{Marziali2017}%
  \BibitemOpen
  \bibfield  {author} {\bibinfo {author} {\bibfnamefont {M.}~\bibnamefont
  {Marziali~Berm\'udez}}, \bibinfo {author} {\bibfnamefont {E.~R.}\
  \bibnamefont {Louden}}, \bibinfo {author} {\bibfnamefont {M.~R.}\
  \bibnamefont {Eskildsen}}, \bibinfo {author} {\bibfnamefont {C.~D.}\
  \bibnamefont {Dewhurst}}, \bibinfo {author} {\bibfnamefont {V.}~\bibnamefont
  {Bekeris}}, \ and\ \bibinfo {author} {\bibfnamefont {G.}~\bibnamefont
  {Pasquini}},\ }\bibfield  {title} {\emph {\bibinfo {title} {{Metastability
  and hysteretic vortex pinning near the order-disorder transition in
  ${\mathrm{NbSe}}_{2}$: Interplay between plastic and elastic energy
  barriers}}},\ }\href {\doibase 10.1103/PhysRevB.95.104505} {\bibfield
  {journal} {\bibinfo  {journal} {Physical Review B}\ }\textbf {\bibinfo
  {volume} {95}},\ \bibinfo {pages} {104505} (\bibinfo {year}
  {2017})}\BibitemShut {NoStop}%
\bibitem [{\citenamefont {Guillam\'on~G\'omez}(2009)}]{Guillamon2009-thesis}%
  \BibitemOpen
  \bibfield  {author} {\bibinfo {author} {\bibfnamefont {I.}~\bibnamefont
  {Guillam\'on~G\'omez}},\ }\emph {\bibinfo {title} {{Orden y desorden en
  superconductividad}}},\ \href@noop {} {Ph.D. thesis},\ \bibinfo  {school}
  {Universidad Aut\'onoma de Madrid} (\bibinfo {year} {2009})\BibitemShut
  {NoStop}%
\bibitem [{\citenamefont {Henderson}\ \emph {et~al.}(1998)\citenamefont
  {Henderson}, \citenamefont {Andrei},\ and\ \citenamefont
  {Higgins}}]{Henderson1998}%
  \BibitemOpen
  \bibfield  {author} {\bibinfo {author} {\bibfnamefont {W.}~\bibnamefont
  {Henderson}}, \bibinfo {author} {\bibfnamefont {E.~Y.}\ \bibnamefont
  {Andrei}}, \ and\ \bibinfo {author} {\bibfnamefont {M.~J.}\ \bibnamefont
  {Higgins}},\ }\bibfield  {title} {\emph {\bibinfo {title} {Plastic motion of
  a vortex lattice driven by alternating current}},\ }\href {\doibase
  10.1103/PhysRevLett.81.2352} {\bibfield  {journal} {\bibinfo  {journal}
  {Phys. Rev. Lett.}\ }\textbf {\bibinfo {volume} {81}},\ \bibinfo {pages}
  {2352} (\bibinfo {year} {1998})}\BibitemShut {NoStop}%
\bibitem [{\citenamefont {Xiao}\ \emph {et~al.}(2004)\citenamefont {Xiao},
  \citenamefont {Dogru}, \citenamefont {Andrei}, \citenamefont {Shuk},\ and\
  \citenamefont {Greenblatt}}]{Xiao2004}%
  \BibitemOpen
  \bibfield  {author} {\bibinfo {author} {\bibfnamefont {Z.~L.}\ \bibnamefont
  {Xiao}}, \bibinfo {author} {\bibfnamefont {O.}~\bibnamefont {Dogru}},
  \bibinfo {author} {\bibfnamefont {E.~Y.}\ \bibnamefont {Andrei}}, \bibinfo
  {author} {\bibfnamefont {P.}~\bibnamefont {Shuk}}, \ and\ \bibinfo {author}
  {\bibfnamefont {M.}~\bibnamefont {Greenblatt}},\ }\bibfield  {title} {\emph
  {\bibinfo {title} {Observation of the vortex lattice spinodal in
  ${\mathrm{nbse}}_{2}$}},\ }\href {\doibase 10.1103/PhysRevLett.92.227004}
  {\bibfield  {journal} {\bibinfo  {journal} {Phys. Rev. Lett.}\ }\textbf
  {\bibinfo {volume} {92}},\ \bibinfo {pages} {227004} (\bibinfo {year}
  {2004})}\BibitemShut {NoStop}%
\bibitem [{\citenamefont {Li}\ \emph {et~al.}(2006)\citenamefont {Li},
  \citenamefont {Andrei}, \citenamefont {Xiao}, \citenamefont {Shuk},\ and\
  \citenamefont {Greenblatt}}]{Li2006}%
  \BibitemOpen
  \bibfield  {author} {\bibinfo {author} {\bibfnamefont {G.}~\bibnamefont
  {Li}}, \bibinfo {author} {\bibfnamefont {E.~Y.}\ \bibnamefont {Andrei}},
  \bibinfo {author} {\bibfnamefont {Z.~L.}\ \bibnamefont {Xiao}}, \bibinfo
  {author} {\bibfnamefont {P.}~\bibnamefont {Shuk}}, \ and\ \bibinfo {author}
  {\bibfnamefont {M.}~\bibnamefont {Greenblatt}},\ }\bibfield  {title} {\emph
  {\bibinfo {title} {Onset of motion and dynamic reordering of a vortex
  lattice}},\ }\href {\doibase 10.1103/PhysRevLett.96.017009} {\bibfield
  {journal} {\bibinfo  {journal} {Phys. Rev. Lett.}\ }\textbf {\bibinfo
  {volume} {96}},\ \bibinfo {pages} {017009} (\bibinfo {year}
  {2006})}\BibitemShut {NoStop}%
\bibitem [{\citenamefont {Yaron}\ \emph {et~al.}(1995)\citenamefont {Yaron},
  \citenamefont {Gammel}, \citenamefont {Huse}, \citenamefont {Kleiman},
  \citenamefont {Oglesby}, \citenamefont {Bucher}, \citenamefont {Batlogg},
  \citenamefont {Bishop}, \citenamefont {Mortensen},\ and\ \citenamefont
  {Clausen}}]{Yaron1995}%
  \BibitemOpen
  \bibfield  {author} {\bibinfo {author} {\bibfnamefont {U.}~\bibnamefont
  {Yaron}}, \bibinfo {author} {\bibfnamefont {P.~L.}\ \bibnamefont {Gammel}},
  \bibinfo {author} {\bibfnamefont {D.~A.}\ \bibnamefont {Huse}}, \bibinfo
  {author} {\bibfnamefont {R.~N.}\ \bibnamefont {Kleiman}}, \bibinfo {author}
  {\bibfnamefont {C.~S.}\ \bibnamefont {Oglesby}}, \bibinfo {author}
  {\bibfnamefont {E.}~\bibnamefont {Bucher}}, \bibinfo {author} {\bibfnamefont
  {B.}~\bibnamefont {Batlogg}}, \bibinfo {author} {\bibfnamefont {D.~J.}\
  \bibnamefont {Bishop}}, \bibinfo {author} {\bibfnamefont {K.}~\bibnamefont
  {Mortensen}}, \ and\ \bibinfo {author} {\bibfnamefont {K.~N.}\ \bibnamefont
  {Clausen}},\ }\bibfield  {title} {\emph {\bibinfo {title} {Structural
  evidence for a two-step process in the depinning of the superconducting
  flux-line lattice}},\ }\href@noop {} {\bibfield  {journal} {\bibinfo
  {journal} {Nature (London)}\ }\textbf {\bibinfo {volume} {376}},\ \bibinfo
  {pages} {753} (\bibinfo {year} {1995})}\BibitemShut {NoStop}%
\bibitem [{\citenamefont {Hess}\ \emph {et~al.}(1989)\citenamefont {Hess},
  \citenamefont {Robinson}, \citenamefont {Dynes}, \citenamefont {Valles},\
  and\ \citenamefont {Waszczak}}]{Hess1989}%
  \BibitemOpen
  \bibfield  {author} {\bibinfo {author} {\bibfnamefont {H.~F.}\ \bibnamefont
  {Hess}}, \bibinfo {author} {\bibfnamefont {R.~B.}\ \bibnamefont {Robinson}},
  \bibinfo {author} {\bibfnamefont {R.~C.}\ \bibnamefont {Dynes}}, \bibinfo
  {author} {\bibfnamefont {J.~M.}\ \bibnamefont {Valles}}, \ and\ \bibinfo
  {author} {\bibfnamefont {J.~V.}\ \bibnamefont {Waszczak}},\ }\bibfield
  {title} {\emph {\bibinfo {title} {{Scanning-Tunneling-Microscope Observation
  of the Abrikosov Flux Lattice and the Density of States near and inside a
  Fluxoid}}},\ }\href {\doibase 10.1103/PhysRevLett.62.214} {\bibfield
  {journal} {\bibinfo  {journal} {Physical Review Letters}\ }\textbf {\bibinfo
  {volume} {62}},\ \bibinfo {pages} {214} (\bibinfo {year} {1989})}\BibitemShut
  {NoStop}%
\bibitem [{\citenamefont {Raes}\ \emph {et~al.}(2014)\citenamefont {Raes},
  \citenamefont {de~Souza~Silva}, \citenamefont {Silhanek}, \citenamefont
  {Cabral}, \citenamefont {Moshchalkov},\ and\ \citenamefont {Van~de
  Vondel}}]{Raes2014}%
  \BibitemOpen
  \bibfield  {author} {\bibinfo {author} {\bibfnamefont {B.}~\bibnamefont
  {Raes}}, \bibinfo {author} {\bibfnamefont {C.~C.}\ \bibnamefont
  {de~Souza~Silva}}, \bibinfo {author} {\bibfnamefont {A.~V.}\ \bibnamefont
  {Silhanek}}, \bibinfo {author} {\bibfnamefont {L.~R.~E.}\ \bibnamefont
  {Cabral}}, \bibinfo {author} {\bibfnamefont {V.~V.}\ \bibnamefont
  {Moshchalkov}}, \ and\ \bibinfo {author} {\bibfnamefont {J.}~\bibnamefont
  {Van~de Vondel}},\ }\bibfield  {title} {\emph {\bibinfo {title} {{Closer look
  at the low-frequency dynamics of vortex matter using scanning susceptibility
  microscopy}}},\ }\href {\doibase 10.1103/PhysRevB.90.134508} {\bibfield
  {journal} {\bibinfo  {journal} {Physical Review B}\ }\textbf {\bibinfo
  {volume} {90}},\ \bibinfo {pages} {134508} (\bibinfo {year}
  {2014})}\BibitemShut {NoStop}%
\bibitem [{\citenamefont {Timmermans}\ \emph {et~al.}(2014)\citenamefont
  {Timmermans}, \citenamefont {Samuely}, \citenamefont {Raes}, \citenamefont
  {Van~de Vondel},\ and\ \citenamefont {Moshchalkov}}]{Timmermans2014}%
  \BibitemOpen
  \bibfield  {author} {\bibinfo {author} {\bibfnamefont {M.}~\bibnamefont
  {Timmermans}}, \bibinfo {author} {\bibfnamefont {T.}~\bibnamefont {Samuely}},
  \bibinfo {author} {\bibfnamefont {B.}~\bibnamefont {Raes}}, \bibinfo {author}
  {\bibfnamefont {J.}~\bibnamefont {Van~de Vondel}}, \ and\ \bibinfo {author}
  {\bibfnamefont {V.~V.}\ \bibnamefont {Moshchalkov}},\ }\bibfield  {title}
  {\emph {\bibinfo {title} {{Dynamic Visualization of Nanoscale Vortex
  Orbits}}},\ }\href {\doibase 10.1021/nn4065007} {\bibfield  {journal}
  {\bibinfo  {journal} {ACS Nano}\ }\textbf {\bibinfo {volume} {8}},\ \bibinfo
  {pages} {2782} (\bibinfo {year} {2014})}\BibitemShut {NoStop}%
\bibitem [{\citenamefont {Guillamon}\ \emph {et~al.}(2008)\citenamefont
  {Guillamon}, \citenamefont {Suderow}, \citenamefont {Guinea},\ and\
  \citenamefont {Vieira}}]{Guillamon2008b}%
  \BibitemOpen
  \bibfield  {author} {\bibinfo {author} {\bibfnamefont {I.}~\bibnamefont
  {Guillamon}}, \bibinfo {author} {\bibfnamefont {H.}~\bibnamefont {Suderow}},
  \bibinfo {author} {\bibfnamefont {F.}~\bibnamefont {Guinea}}, \ and\ \bibinfo
  {author} {\bibfnamefont {S.}~\bibnamefont {Vieira}},\ }\bibfield  {title}
  {\emph {\bibinfo {title} {{Intrinsic atomic-scale modulations of the
  superconducting gap of 2H-$\mathrm{NbSe}_{2}$}}},\ }\href {\doibase
  10.1103/PhysRevB.77.134505} {\bibfield  {journal} {\bibinfo  {journal}
  {Physical Review B}\ }\textbf {\bibinfo {volume} {77}},\ \bibinfo {pages}
  {134505} (\bibinfo {year} {2008})}\BibitemShut {NoStop}%
\bibitem [{\citenamefont {Galvis}\ \emph {et~al.}(2017)\citenamefont {Galvis},
  \citenamefont {Herrera}, \citenamefont {Guillam{\'o}n}, \citenamefont
  {Vieira},\ and\ \citenamefont {Suderow}}]{Galvis2017}%
  \BibitemOpen
  \bibfield  {author} {\bibinfo {author} {\bibfnamefont {J.}~\bibnamefont
  {Galvis}}, \bibinfo {author} {\bibfnamefont {E.}~\bibnamefont {Herrera}},
  \bibinfo {author} {\bibfnamefont {I.}~\bibnamefont {Guillam{\'o}n}}, \bibinfo
  {author} {\bibfnamefont {S.}~\bibnamefont {Vieira}}, \ and\ \bibinfo {author}
  {\bibfnamefont {H.}~\bibnamefont {Suderow}},\ }\bibfield  {title} {\emph
  {\bibinfo {title} {{Vortex cores and vortex motion in superconductors with
  anisotropic Fermi surfaces}}},\ }\href {\doibase
  https://doi.org/10.1016/j.physc.2016.07.023} {\bibfield  {journal} {\bibinfo
  {journal} {Physica C: Superconductivity and its Applications}\ }\textbf
  {\bibinfo {volume} {533}},\ \bibinfo {pages} {2} (\bibinfo {year}
  {2017})}\BibitemShut {NoStop}%
\bibitem [{\citenamefont {Embon}\ \emph {et~al.}(2015)\citenamefont {Embon},
  \citenamefont {Anahory}, \citenamefont {Suhov}, \citenamefont {Halbertal},
  \citenamefont {Cuppens}, \citenamefont {Yakovenko}, \citenamefont {Uri},
  \citenamefont {Myasoedov}, \citenamefont {Rappaport}, \citenamefont {Huber},
  \citenamefont {Gurevich},\ and\ \citenamefont {Zeldov}}]{Embon2015}%
  \BibitemOpen
  \bibfield  {author} {\bibinfo {author} {\bibfnamefont {L.}~\bibnamefont
  {Embon}}, \bibinfo {author} {\bibfnamefont {Y.}~\bibnamefont {Anahory}},
  \bibinfo {author} {\bibfnamefont {A.}~\bibnamefont {Suhov}}, \bibinfo
  {author} {\bibfnamefont {D.}~\bibnamefont {Halbertal}}, \bibinfo {author}
  {\bibfnamefont {J.}~\bibnamefont {Cuppens}}, \bibinfo {author} {\bibfnamefont
  {A.}~\bibnamefont {Yakovenko}}, \bibinfo {author} {\bibfnamefont
  {A.}~\bibnamefont {Uri}}, \bibinfo {author} {\bibfnamefont {Y.}~\bibnamefont
  {Myasoedov}}, \bibinfo {author} {\bibfnamefont {M.~L.}\ \bibnamefont
  {Rappaport}}, \bibinfo {author} {\bibfnamefont {M.~E.}\ \bibnamefont
  {Huber}}, \bibinfo {author} {\bibfnamefont {A.}~\bibnamefont {Gurevich}}, \
  and\ \bibinfo {author} {\bibfnamefont {E.}~\bibnamefont {Zeldov}},\
  }\bibfield  {title} {\emph {\bibinfo {title} {{Probing dynamics and pinning
  of single vortices in superconductors at nanometer scales}}},\ }\href@noop {}
  {\bibfield  {journal} {\bibinfo  {journal} {Scientific Reports}\ }\textbf
  {\bibinfo {volume} {5}},\ \bibinfo {pages} {7598} (\bibinfo {year}
  {2015})}\BibitemShut {NoStop}%
\bibitem [{\citenamefont {Campbell}(1969)}]{Campbell1969}%
  \BibitemOpen
  \bibfield  {author} {\bibinfo {author} {\bibfnamefont {A.~M.}\ \bibnamefont
  {Campbell}},\ }\bibfield  {title} {\emph {\bibinfo {title} {{The response of
  pinned flux vortices to low-frequency fields}}},\ }\href {\doibase
  10.1088/0022-3719/2/8/318} {\bibfield  {journal} {\bibinfo  {journal}
  {Journal of Physics C: Solid State Physics}\ }\textbf {\bibinfo {volume}
  {2}},\ \bibinfo {pages} {1492} (\bibinfo {year} {1969})}\BibitemShut
  {NoStop}%
\bibitem [{\citenamefont {Campbell}(1971)}]{Campbell1971}%
  \BibitemOpen
  \bibfield  {author} {\bibinfo {author} {\bibfnamefont {A.~M.}\ \bibnamefont
  {Campbell}},\ }\bibfield  {title} {\emph {\bibinfo {title} {{The interaction
  distance between flux lines and pinning centres}}},\ }\href {\doibase
  10.1088/0022-3719/4/18/023} {\bibfield  {journal} {\bibinfo  {journal}
  {Journal of Physics C: Solid State Physics}\ }\textbf {\bibinfo {volume}
  {4}},\ \bibinfo {pages} {3186} (\bibinfo {year} {1971})}\BibitemShut
  {NoStop}%
\bibitem [{\citenamefont {Bean}(1962)}]{Bean1962}%
  \BibitemOpen
  \bibfield  {author} {\bibinfo {author} {\bibfnamefont {C.~P.}\ \bibnamefont
  {Bean}},\ }\bibfield  {title} {\emph {\bibinfo {title} {{Magnetization of
  hard superconductors}}},\ }\href {\doibase 10.1103/PhysRevLett.8.250}
  {\bibfield  {journal} {\bibinfo  {journal} {Physical Review Letters}\
  }\textbf {\bibinfo {volume} {8}},\ \bibinfo {pages} {250} (\bibinfo {year}
  {1962})}\BibitemShut {NoStop}%
\bibitem [{\citenamefont {van~der Beek}\ \emph {et~al.}(1993)\citenamefont
  {van~der Beek}, \citenamefont {Geshkenbein},\ and\ \citenamefont
  {Vinokur}}]{vanderBeek1993}%
  \BibitemOpen
  \bibfield  {author} {\bibinfo {author} {\bibfnamefont {C.~J.}\ \bibnamefont
  {van~der Beek}}, \bibinfo {author} {\bibfnamefont {V.~B.}\ \bibnamefont
  {Geshkenbein}}, \ and\ \bibinfo {author} {\bibfnamefont {V.~M.}\ \bibnamefont
  {Vinokur}},\ }\bibfield  {title} {\emph {\bibinfo {title} {{Linear and
  nonlinear ac response in the superconducting mixed state}}},\ }\href
  {\doibase 10.1103/PhysRevB.48.3393} {\bibfield  {journal} {\bibinfo
  {journal} {Physical Review B}\ }\textbf {\bibinfo {volume} {48}},\ \bibinfo
  {pages} {3393 } (\bibinfo {year} {1993})}\BibitemShut {NoStop}%
\bibitem [{\citenamefont {Brandt}(1994)}]{Brandt1994a}%
  \BibitemOpen
  \bibfield  {author} {\bibinfo {author} {\bibfnamefont {E.~H.}\ \bibnamefont
  {Brandt}},\ }\bibfield  {title} {\emph {\bibinfo {title} {{Thin
  superconductors in a perpendicular magnetic ac field: General formulation and
  strip geometry}}},\ }\href {\doibase 10.1103/PhysRevB.49.9024} {\bibfield
  {journal} {\bibinfo  {journal} {Physical Review B}\ }\textbf {\bibinfo
  {volume} {49}},\ \bibinfo {pages} {9024} (\bibinfo {year}
  {1994})}\BibitemShut {NoStop}%
\bibitem [{\citenamefont {Valenzuela}\ and\ \citenamefont
  {Bekeris}(2000)}]{Valenzuela2000}%
  \BibitemOpen
  \bibfield  {author} {\bibinfo {author} {\bibfnamefont {S.~O.}\ \bibnamefont
  {Valenzuela}}\ and\ \bibinfo {author} {\bibfnamefont {V.}~\bibnamefont
  {Bekeris}},\ }\bibfield  {title} {\emph {\bibinfo {title} {Plasticity and
  memory effects in the vortex solid phase of twinned
  ${\mathrm{yba}}_{2}{\mathrm{cu}}_{3}{O}_{7}$ single crystals}},\ }\href
  {\doibase 10.1103/PhysRevLett.84.4200} {\bibfield  {journal} {\bibinfo
  {journal} {Phys. Rev. Lett.}\ }\textbf {\bibinfo {volume} {84}},\ \bibinfo
  {pages} {4200} (\bibinfo {year} {2000})}\BibitemShut {NoStop}%
\bibitem [{\citenamefont {Stamopoulos}\ \emph {et~al.}(2002)\citenamefont
  {Stamopoulos}, \citenamefont {Pissas},\ and\ \citenamefont
  {Bondarenko}}]{Stamopoulos2002}%
  \BibitemOpen
  \bibfield  {author} {\bibinfo {author} {\bibfnamefont {D.}~\bibnamefont
  {Stamopoulos}}, \bibinfo {author} {\bibfnamefont {M.}~\bibnamefont {Pissas}},
  \ and\ \bibinfo {author} {\bibfnamefont {A.}~\bibnamefont {Bondarenko}},\
  }\bibfield  {title} {\emph {\bibinfo {title} {Possible reordering of vortex
  matter near the end point of the second peak line in the
  ${\mathrm{yba}}_{2}{\mathrm{cu}}_{3}{\mathrm{o}}_{7\ensuremath{-}\ensuremath{\delta}}$
  compound}},\ }\href {\doibase 10.1103/PhysRevB.66.214521} {\bibfield
  {journal} {\bibinfo  {journal} {Phys. Rev. B}\ }\textbf {\bibinfo {volume}
  {66}},\ \bibinfo {pages} {214521} (\bibinfo {year} {2002})}\BibitemShut
  {NoStop}%
\bibitem [{Mar()}]{Marziali2015sup}%
  \BibitemOpen
  \href@noop {} {\emph {\bibinfo {title} {{\emph{Supplementary material of
  Ref.\ [\onlinecite{Marziali2015}]}}}}}\BibitemShut {NoStop}%
\bibitem [{\citenamefont {Chandra~Ganguli}\ \emph {et~al.}(2015)\citenamefont
  {Chandra~Ganguli}, \citenamefont {Singh}, \citenamefont {Saraswat},
  \citenamefont {Ganguly}, \citenamefont {Bagwe}, \citenamefont {Shirage},
  \citenamefont {Thamizhavel},\ and\ \citenamefont
  {Raychaudhuri}}]{ChandraGanguli2015}%
  \BibitemOpen
  \bibfield  {author} {\bibinfo {author} {\bibfnamefont {S.}~\bibnamefont
  {Chandra~Ganguli}}, \bibinfo {author} {\bibfnamefont {H.}~\bibnamefont
  {Singh}}, \bibinfo {author} {\bibfnamefont {G.}~\bibnamefont {Saraswat}},
  \bibinfo {author} {\bibfnamefont {R.}~\bibnamefont {Ganguly}}, \bibinfo
  {author} {\bibfnamefont {V.}~\bibnamefont {Bagwe}}, \bibinfo {author}
  {\bibfnamefont {P.}~\bibnamefont {Shirage}}, \bibinfo {author} {\bibfnamefont
  {A.}~\bibnamefont {Thamizhavel}}, \ and\ \bibinfo {author} {\bibfnamefont
  {P.}~\bibnamefont {Raychaudhuri}},\ }\bibfield  {title} {\emph {\bibinfo
  {title} {Disordering of the vortex lattice through successive destruction of
  positional and orientational order in a weakly pinned co0.0075nbse2 single
  crystal}},\ }\href {http://dx.doi.org/10.1038/srep10613} {\bibfield
  {journal} {\bibinfo  {journal} {Scientific Reports}\ }\textbf {\bibinfo
  {volume} {5}},\ \bibinfo {pages} {10613} (\bibinfo {year}
  {2015})}\BibitemShut {NoStop}%
\end{thebibliography}%

\end{document}